\documentclass[prx,reprint,twocolumn,showpacs,superscriptaddress]{revtex4-2}
\usepackage[export]{adjustbox}
\usepackage{amsmath,amssymb,bm,mathrsfs,graphicx, braket, times, mathtools}
\usepackage[colorlinks=true,citecolor=blue,linkcolor=blue]{hyperref}
\usepackage{multirow}
\usepackage[all,cmtip]{xy}
\usepackage[normalem]{ulem}
\usepackage[usenames,dvipsnames]{color}
\usepackage{tikz-cd}
\usepackage[compat=1.0.0]{tikz-feynman}
\usepackage{bbding}
\usepackage{graphicx}
\usepackage{subfigure}
\usepackage{bbm}
\usepackage{multirow}

\usepackage[utf8]{inputenc}
\usepackage[english]{babel}
\usepackage{amsthm}
\usepackage{wasysym}

\renewcommand{\vec}[1]{\mathbf{#1}}

 \newcommand{\Tr}{\text{Tr}}
 \newcommand{\ev}[1]{\langle #1\rangle}
 \newcommand{\ga}{\hat \gamma}

 \newcommand{\hc}{\hat c}
 \newcommand{\hd}{\hat d}
 \newcommand{\id}{\mathbbm{1}}
 
 \newcommand{\ep}{\varepsilon}
 
 \newcommand{\mH}{\mathcal{H}}
 \newcommand{\tvec}[1]{\tilde{\vec{#1}}}

\usepackage[normalem]{ulem}

\begin{document}
\title{Constructive Fermionic Matrix Product States for Projected Fermi Sea}

\author{Kangle Li}
\thanks{These authors contributed equally.}
\affiliation{Department of Physics, Hong Kong University of Science and Technology, Clear Water Bay, Hong Kong SAR, China}

\author{Yan-Bai Zhang}
\thanks{These authors contributed equally.}
\affiliation{Department of Physics, Hong Kong University of Science and Technology, Clear Water Bay, Hong Kong SAR, China}

\author{Hoi Chun Po}
\email{hcpo@ust.hk}
\affiliation{Department of Physics, Hong Kong University of Science and Technology, Clear Water Bay, Hong Kong SAR, China}

\begin{abstract}
Projected wave functions offer a means for incorporating local correlation effects in gapless electronic phases of matter like metals. 
Although such wave functions can be readily specified formally, it is challenging to compute their associated physical observables.
Tensor network approaches offer a modern numerical method for this task. 
In this work, we develop and demonstrate a constructive tensor-network approach for obtaining physical quantities, like fermion two-point functions and density-density correlation functions, for one-dimensional projected Fermi sea states. 
We benchmark our method against exact analytical results for spin-$1/2$ electrons subjected to the Gutzwiller projection, and then present results on spinless fermions with nearest-neighbor repulsion. For a state with two pairs of Fermi points, we reveal a correlation-tuning of the characteristic wave vector of the charge density modulation in the system.
\end{abstract}

\maketitle

\section{Introduction}
Projected wave functions, like the Gutzwiller ansatz~\cite{gutzwiller1963effect,PhysRev.137.A1726_gutzwiller_1965}, provide a general framework for incorporating electron correlation effects into the description of gapless phases of matter~\cite{PhysRevB.2.4302_Brinkman_Rice,RevModPhys.56.99_Vollhardt,zhang1988renormalised,vollhardt1990gutzwiller,PhysRevLett.59.121_Metzner1987,edegger2007gutzwiller,deng2008lda+,PhysRevB.79.165114_XDai,PhysRevB.79.075114_XDai,bunemann2009gutzwiller_variation,cheng2021foundations,cheng2021variational,cheng2022precise,cheng2023gauge,orus2014practical,orus2019tensor}). Intuitively, such wave functions contain two key ingredients: a ``projection,'' which embodies local correlation effects like the energy penalty of double electron occupancy on the same site, and a pre-projection parent state, like a filled Fermi sea, which captures the qualitative features of the low-energy physical properties of interest.
In strongly correlated systems, projected wave functions also provide a way for constructing fractionalized ground states by projecting out unphysical states in an enlarged parton Hilbert space~\cite{liu2012gutzwiller_tu,wu2020tensor_mpo_mps_tu,petrica2021finite,budaraju2024simulating}.

Valuable insights into the interplay between local correlation effects and the characteristic features anticipated from a mean-field starting point can be obtained by investigating how the projection strength affects physical observables.
Generally speaking, however, such physical observables are challenging to compute. While exact analytical results are available for certain special models~\cite{gebhard1988correlation,metzner1988analytic,kollar2002exact,PhysRevLett.59.121_Metzner1987,vollhardt1990gutzwiller}, in general either approximate perturbative treatment or numerical Monte Carlo methods are needed for extracting the physical properties of a projected wave function~\cite{edegger2007gutzwiller,deng2008lda+,PhysRevB.79.165114_XDai,PhysRevB.79.075114_XDai}. 

Tensor network (TN) states provide an alternative starting point for describing strongly correlated phases of matter~\cite{RevModPhys.77.259_dmrg,schollwock2011density_dmrg,verstraete2004renormalization_peps,perez2006matrix,verstraete2008matrix,PhysRevB.78.205116_gu_TERG,ORUS2014117_review,cirac2009renormalization,fishman2015compression,PhysRevA.104.012415_Plenio}. 
More broadly, TN also serves as a natural framework for describing a projected wave function: once the TN form of the pre-projection parent state is obtained, the projection can be implemented readily by applying additional local operators to the physical legs of the tensors. Earlier works have demonstrated the promise of this approach in both gapped and gapless systems in which double occupancy of the electrons or partons are projected away~\cite{wu2020tensor_mpo_mps_tu,jin2021density_tu,jin2020efficient_tu,yang2023projected_tu, li2023u_tu,budaraju2024simulating, petrica2021finite}. Aside from providing a new practical means to evaluate physical observables in Gutzwiller wave functions, such projected states could also be used for initializing variational TN calculations in which the competitiveness of a more entangled states \cite{jin2021density_tu, petrica2021finite}, like those featuring possible Fermi surfaces, could be explored.

In this work, we develop and demonstrate a constructive approach for the TN-aided evaluation of projected wave functions. We restrict our scope to projected Fermi sea states in one spatial dimension, which allows us to leverage the highly efficient matrix product state (MPS) and transfer matrix frameworks for evaluating physical observables. 
We report results on the fermion two-point functions, as well as the density-density and spin-spin correlation functions, of projected Fermi sea states. 
Our results are first benchmarked against the analytical results available in the standard Gutzwiller problem of a filled Fermi sea of spin-$1/2$  electrons. 
Next, we apply our method to study spinless fermions subjected to nearest-neighbor repulsion, for which exact analytical results are not available. 
This allows us to explore the interplay between such local correlations and the fermiology of the mean-field parent state. 
When there are two pairs of distinct Fermi points, we reveal a competition between the parent-state fermiology and the local correlation effects in the charge density modulation. 
In contrast to expectations from the weak coupling limit, we find that a pronounced peak in the static structure factor develops at a wave vector which cannot be simply inferred from the Fermi wave vectors characterizing the Fermi surfaces. 

Our approach starts from constructing the MPS representation of a free-fermionic state at the level of correlation matrix~\cite{kraus2010fermionic_fGPEPS,fishman2015compression,dubail2015tensor,schuch_bauer_2019matrix,petrica2021finite,Jin_Hatree-Fock-Bogoliubov,li2023u_tu}, which encodes all the data present in the free-fermion state. 
Various technical approaches exist for this task, including the successive discovery and application of local rotations~\cite{fishman2015compression}, as well as the variational optimization of ansatz local tensors~\cite{schuch_bauer_2019matrix,perez2006matrix,schollwock2011density_dmrg}.
Here, we construct fermionic Gaussian MPS based on Schmidt decomposition. This can be achieved either by sweeping across the whole finite chain directly 
\cite{orus2014practical}, or by first decomposing the locally defined Wannier functions of the target state and then stacking them into the full MPS representation~\cite{he2024stacked}.
The local Gaussian tensors can then be turned into many-body ones~\cite{PhysRevA.104.012415_Plenio}, which enables the subsequent incorporation of Gutzwiller-like projections. 
In our approach, the fermionic nature of the tensors is maintained and the many-body bond spaces can all be interpreted as fermionic Fock spaces. This facilitates the translation of Gaussian-level tensor contractions to the many-body treatment.

Although the scope of this work is restricted to projected Fermi sea states in one spatial dimension, we note that our general constructive approach for projected wave functions can be employed in any spatial dimensions, with the main difficulty stemming only form the usual increased complexity in evaluating physical observables in higher-dimensional TN states. Our work helps pave the way for exploring the power of TN methods in studying gapless phases of fermions in higher dimensions.

The paper is organized as follows. First we introduce the basic tools needed for construction and computation of fermionic MPS in Section \ref{sec:basic_tools}. Then we show the explicit example of 1d Fermi sea state with two schemes in Section \ref{sec:construct_fgmps}. Gutzwiller projected states are considered in Section \ref{sec:gutz_wave}, where we benchmark with some known analytical results and present our main results on the interplay between nearest-neighbor repulsion and the fermiology of the parent state. 

\section{Basic Tools for fermionic MPS}\label{sec:basic_tools}

\subsection{Overview of the procedures}

Our starting point is a free-fermion state corresponding to a filled Fermi sea state. All the data in this parent state are encapsulated in its (static) two-point functions. 
On a lattice, these two-point functions are typically viewed as entries of a matrix, called the covariance matrix~\cite{kraus2010fermionic_fGPEPS,bravyi2004lagrangian,peschel2012entanglement}
. If the state has a definite particle number, a more restricted form, defined below as the correlation matrix, would suffice for describing the state.
Since we are interested in projected Fermi sea states, we assume the particle number of the state is conserved throughout. Correspondingly, we specialize our discussion to the case of correlation matrices, although the same operations can be readily generalized to covariance matrices. We also use ``free-fermion'' and ``Gaussian'' interchangeably, with the two understood to be the same when the particle number is conserved.

Next, we represent the parent state as a fermionic Gaussian MPS (fGMPS), for which all local tensors are represented by their correlation matrices\cite{fishman2015compression,petrica2021finite,Jin_Hatree-Fock-Bogoliubov,PhysRevA.104.012415_Plenio}. 
A key step in obtaining the desired MPS representation, as in the usual case, is the Schmidt decomposition of the state with respect to a bipartition of the system.
For Gaussian states, the required Schmidt decompositions can be performed in terms of the correlation matrices (Section~\ref{sec:tool_fGMPS}). 
Conversely, the decomposed pieces can be recombined by performing suitable partial traces, which for Gaussian states can again be achieved through a simple contraction formula (Section~\ref{sec:tool_partial_trace}).

Through an MPS representation of the parent Gaussian state, local correlation effects can now be incorporated readily by applying suitable matrix product operators to the parent state. We will focus on the operators corresponding to soft projectors which either penalize double occupancy of spin-$1/2$ fermions, or the simultaneous presence of spinless fermions on two adjacent sites. As such operators are interacting in nature, we need to first convert all the local tensors in the parent fGMPS into a many-body form. 
This can be achieved through the standard Jordan-Wigner (JW) transformation, which also allows one to compute the contraction of local fermionic tensors through suitably defined matrices (Section ~\ref{sec:tool_many_body}). We can then extract physical observables from the MPS using transfer matrix techniques.

Throughout the paper, we denote $\hc^\dagger$/ $\hc$ as the creation/ annihilation operators of complex fermions. 

\subsection{Fermionic Gaussian matrix product state}\label{sec:tool_fGMPS}
Thanks to the Wick's theorem, all physical information of a free-fermion state is encoded in its correlation matrix $C_{ij} = \ev{\hc_i \hc_j^\dagger}$. We discuss in this subsection how the usual MPS techniques can be adapted for correlation matrices.\\

\noindent {\itshape Schmidt Decomposition}. Given a free-fermion state, the Schmidt decomposition of the state with respect to a subregion $A$ and its complement $B$ can be done using only the data in the correlation matrix $C$. Let us write $C$ in a block-matrix form as 
\begin{equation}
    C = \left(\begin{array}{cc}
        C_A & C_{AB} \\
        C_{AB}^\dagger & C_B 
    \end{array}
    \right).
\end{equation}
We focus on the restricted correlation matrix $C_A$ and diagonalize it with a unitary matrix, i.e. $C_A= U\Lambda U^\dagger$. Then the eigenvalues $\lambda_i$ are related to the Schmidt weights: for $\lambda_i = 0$ ($\lambda_i = 1$), the mode is fully filled (empty), contributing no entanglement across the bipartition\cite{peschel2012entanglement,surace2022fermionic}. The rotated matrix 
\begin{equation}
    C' = \left(\begin{array}{cc}
       \Lambda  & U^\dagger C_{AB} \\
        C_{AB}U & C_B
    \end{array}\right)
\end{equation}
is still a pure state, with the modes corresponding to the block occupied by $\Lambda$ being regarded as bond modes. 
Since modes with $\lambda_i = 0$ or $1$ can only be exactly filled or empty in any Schmidt states with finite Schmidt weights, they can be stripped from the bond Hilbert space. 
In practice, we may set a small numerical threshold $\varepsilon$ such that all modes with $\lambda_i<\ep $ or $ \lambda>1-\ep$ are treated as frozen. These frozen modes in $C'$ can be stripped away  by contracting with the suitable fully filled or empty states.
Alternatively, one can also simply remove the rows and columns corresponding to these modes, and then incorporate an orthonormalization step to maintain the purity of the reduced state (see appendices of \cite{he2024stacked} for more details). 

If the state $C'$ is truncated, the unitary $U$ becomes an isometry. For example, if a mode $\lambda_i\approx 0$ is frozen to filled, then we just remove the corresponding column vector in $U$; but if a mode $\lambda_i\approx 1$ in $C'$ is frozen to empty, then we need to keep it for purification, see below.\\

\noindent {\itshape Kernel purification.} The above Schmidt process returns a unitary $U$ and a correlation matrix $C'$. We can obtain a fermionic Gaussian state corresponding to $U$ by doing purification based on a {\itshape kernel state} $C^{ker}$. 
This procedure can be viewed as a specialized form of the channel-state duality~\cite{jamiolkowski1972linear, choi1975completely}, in which we ensure the purified state is also fermionic Gaussian.
Generally, the kernel state can be chosen to be a maximally entangled state between the physical Hilbert space and an auxiliary Hilbert space of the same dimension.
Here, we use a free-fermion kernel state with the correlation matrix 
\begin{equation}
    C^{ker} = \frac{1}{2}\left( \begin{array}{cc}
        \id & s\id \\
        s^*\id & \id
    \end{array} \right),
\end{equation}
where $s$ is an arbitrary complex phase factor. In terms of creation operators, the state is (up to a normalization factor) 
\begin{equation}\label{eq:kernel}
    \ket{\psi^{ker}(s)} = \prod_{i} (-\hc_i^\dagger + s \hd_i^\dagger)\ket{0}, 
\end{equation}
where we use $\hc^\dagger_i$ to denote the physical modes in $U$ and $\hd^\dagger_i$ to denote the auxiliary modes introduced for purification. 
For simplicity, we choose $s=1$. Then the unitary is purified into a free-fermion state with the correlation matrix 
\begin{equation}
   C_{\ket{U}} = \left(\begin{array}{cc}
       U &  \\
        & \id
   \end{array}\right)C^{ker} \left(\begin{array}{cc}
       U^\dagger &  \\
        & \id
        \end{array} \right).
\end{equation}
If $U$ is an $m\times m$ matrix, then we get a free-fermion state with $2m$ modes, the first $m$ modes are physical modes in subregion $A$ and the second set of $m$ modes are assigned to the bonds. 

As previously mentioned, the unitary matrix $U$ may be truncated to an isometry when there are frozen modes in the Schmidt decomposition. Suppose the isometry, denoted by $U'$, is composed of $n_1$ dynamical modes, assembled into the columns of a submatrix $U^{(1)}$, and another $n_2$ number of modes, corresponding to the submatrix $U^{(2)}$, which have $\lambda=1$ and are therefore frozen. Then, we can purify the isometry by applying 
\begin{equation}
    \tilde U = \left(
    \begin{array}{ccc}
       U^{(1)}_{m\times n_1}  &  & U^{(2)}_{m\times n_2} \\
         & \id_{n_1\times n_1} &
    \end{array}
    \right)
\end{equation}
to a kernel state together with some frozen modes
\begin{equation}
    \tilde C^{ker} = \left(\begin{array}{cc}
        C^{ker}_{2n_1\times 2n_1} &  \\
         & \id_{n_2\times n_2}
    \end{array}\right),
\end{equation}
to get a $(m+n_1)$-mode purified state represented by correlation matrix $C_{\ket{\tilde U}} = \tilde U \tilde C^{ker} \tilde U^\dagger$.

The Schmidt decomposition above is canonical on one side; if $A$ resides on the left of $B$, then the fermionic state $\ket{U}$ is left-canonical as a tensor. Alternatively, one can also perform a left-right-canonical Schmidt decomposition, with two unitary matrices $U$ and $V$ diagonalizing $C_A$ and $C_B$. Then we obtain three pieces 
\begin{equation}
    \begin{split}
    C_{\ket{U}} &= \left( \begin{array}{cc}
        \id & U \\
        U^\dagger & \id  
    \end{array}
    \right), \\
    C_{\Lambda} &= \left( \begin{array}{cc}
        \Lambda_0 & U^\dagger C_{01}V \\
        V^\dagger C_{01}^\dagger U & \Lambda_1  
    \end{array}
    \right), \\
    C_{\ket{V}} &= \left( \begin{array}{cc}
        \id & V^\dagger \\
        V & \id  
    \end{array}
    \right), \\
    \end{split}
\end{equation}
where $C_{\ket{U}}$ is left-canonical, $C_{\ket{V}}$ is right-canonical, and $C_{\Lambda}$ is a ``diagonal'' state.

The Schmidt decomposition serves as a basic step for obtaining a MPS representation of a free-fermion state, as discussed in Sec.\ \ref{sec:construct_fgmps}.

\subsection{Partial Trace and Kernel Contraction}\label{sec:tool_partial_trace}
Having discussed how a fermionic state can be Schmidt-decomposed into local states, we next address how such local states can be ``contracted'' back into a global one. 
The strategy, which applies equally well in the many-body case as is needed later, is to perform a partial trace in the tensor product Hilbert space. 

Consider two density matrices $\hat\rho_{PB}\in\mathcal{H}_{P}\otimes\mathcal{H}_B$ and $\hat\rho'_B\in\mathcal{H}_B$. The partial trace of the two density matrices $ \tilde \rho_P = \Tr_B[\hat \rho_{PB} \hat \rho'_B]$ is another density matrix defined in Hilbert space $\mathcal{H}_P$. 
The mentioned contraction can be rephrased in terms of such partial traces.
Given two density matrices $\hat \rho_{AB}\in \mathcal{H}_A\otimes_f\mathcal{H}_{B}$ and $\hat\rho_{B'C} \in\mathcal{H}_{B'}\otimes_f\mathcal{H}_{C}$, where the two sub-Hilbert spaces $\mathcal{H}_B$ and $\mathcal{H}_{B'}$ has the same dimension, we define the $BB'$ contracted state to be  
\begin{equation}
    \hat \rho_{AC} = \Tr\left[(\hat\rho_{AB}\otimes_f\hat\rho_{B'C})\rho^{ker}_{BB'}\right]
\end{equation}
where $\hat \rho_{BB'}^{ker}$ is the density matrix of a kernel state, which implicitly defines an identification of the fermion modes in Hilbert space $B$ with those in $B'$. We can chose this kernel state to be a pure free-fermion state with correlation matrix taking the same form as in Eq.\ \eqref{eq:kernel}, with the complex phase $\tilde s$. For consistency, however, the phase factor has to be chosen to be $\tilde s = -s^*$ \cite{he2024stacked}, where $s$ denotes the phase chosen for the purification kernel (we choose $s=1$ and hence $\tilde s = -1$). We call this $\hat \rho^{ker}_{BB'}$ the {\itshape contraction kernel}. 

Specializing to free-fermion states, the discussion above also returns a free-fermion state after contraction. 
The partial trace formula for correlation matrices~\cite{bravyi2004lagrangian, kraus2010fermionic_fGPEPS, schuch_bauer_2019matrix, yang2023projected_tu, he2024stacked} is given as follows: denote the two states as $C$ for $\hat\rho_{AB}$ and $\tilde C_{B}$ for $\hat{\tilde{\rho}}_{B}$ with
\[
    C = \left( \begin{array}{cc}
        C_A & C_{AB} \\
        C_{AB}^\dagger & C_B \\ 
    \end{array}\right), 
\]
then the contracted state $C'$ is given by 
\begin{equation}
    C' = C_A - 2C_{AB}\left[2C_B-\id+(2\tilde C_{B}-\id)^{-1}\right]^{-1}C_{AB}^\dagger,
\end{equation}
and if $\tilde C_{B}$ is a pure state, this can be further simplified by $(2\tilde C_{B}-\id )^2=\id$,
\begin{equation}
    C' = C_A -C_{AB}[C_B+ \tilde C_{B}-\id]^{-1}C_{AB}^\dagger.
\end{equation}
For two Gaussian states $C_{AB}, C_{B'C}$ with $B,B'$ to be contracted, our kernel contraction formula is obtained by replacing the $C$ by $C_{AB}\oplus C_{B'C}$, and replacing $\tilde C_{B}$ by $C^{ker}$. We also remark that the kernel contraction of $\hat\rho_{AB}$ with a state $\hat\rho_{B'}$ is equivalent to the partial trace of $\hat\rho_{AB}$ and the particle-hole dual of $\hat\rho_{B'}$, which has correlation matrix $\id-C_{B'}$. 

\subsection{Many-Body fermionic matrix product states}\label{sec:tool_many_body}
To move on to correlated states, we promote the fGMPS to a many-body representation.
Given a set of pure Gaussian states as local tensors of fMPS, we can get many-body local tensors by writing down the wave functions of each Gaussian states. Starting from the correlation matrix, this can be achieved by taking the Slater determinant of all the filled modes. Equivalently, in second-quantized form we can write
\begin{equation}
    |\Psi \rangle = \hat c^\dagger_{f_0}\hat c^\dagger_{f_1}\cdots \hat c^\dagger_{f_{\nu-1}} | {\rm vac} \rangle,
\end{equation}
where $| {\rm vac} \rangle$ denotes the fermionic vacuum state, $\nu$ denotes the charge of the state, and $\hat c^\dagger_{f_i}$ for $i=0,1,\dots, \nu-1$ are the filled modes.
For numerical calculations, we may represent all the locally defined fermionic operators using a JW transformation \footnote{We use a JW convention such that, taking four fermion modes as an example, the many-body states are ordered according to $\ket{0000}, \ket{0001}, \ket{0010},...$}, and with that $|\Psi \rangle$ can be represented as a $2^N$ dimensional vector, where $N \geq \nu$ denotes the number of fermion modes in the local Hilbert space.

As previously discussed, the kernel contraction of local tensors is defined through the partial trace. Since all local tensors are viewed as pure fermionic states, this partial trace process can be done in terms of the wave functions directly. Suppose the wave functions of two local states are represented numerically by two vectors $\Phi_{AB}$ and $\Psi_{B'C}$ in occupation number basis. The contracted state is given by
\begin{equation}\label{eq:wavefunction_contraction}
    ([\hat P_{A}])^{N_B} \Phi_{AB}[K]_{BB'} \Psi_{B'C},
\end{equation}
where $[\hat P]_C$ is the matrix form of a parity operator of the modes in sub-Hilbert space $\mH_A$ and its power $N_B$ is the number of modes in $\mH_B$. This extra parity operator is due to the JW representation and does not effect if $N_B$ is even (see Appendix~\ref{appendix:transfer_mat_and_computation} for a derivation).

The kernel purification can be extended to the many-body setting. Given a many-body operator $\hat O$, the kernel purification is $\ket{\hat O} = (\hat O\otimes \id) \ket{\psi^{ker}}$. This can be reconciled with the thermofield double state if $\hat O$ is the square root of a density matrix. This purification preserves fermionic structure naturally, and the operation of $\hat O$ to a quantum state $\ket{\varphi}$ is equal to the contraction of purified $\ket{\hat O}$ and $\varphi$, up to an overall sign which depends on the parity of the state (Appendix \ref{appendix:transfer_mat_and_computation}). 
This allows us to treat fermionic matrix product operators in terms of local fermionic states as well. \\

\noindent {\itshape Computing Observables}. Transfer matrix is used to compute observables. After tracing out the physical modes of a local tensor and regrouping elements, we are left with a transfer matrix $M_i$ whose rows/ columns are corresponding to the left/ right bonds. Combining with contraction kernels they can be multiplied together, and the expectation value is the trace of the product 
\begin{equation}\label{eq:trans_mat_exp_values}
    \ev{\hat O_i } = \frac{\Tr\left[M_0^{(K)} M_1^{(K)} ... M_{\hat O_i}^{(K)} M_{i+1}^{(K)} ... M_{L-1}^{(K)} \Xi^{N^{\vec t}+1} \right]}{\Tr\left[\prod_i M_i^{(K) } \Xi^{N^{\vec t}+1} \right]},
\end{equation}
where $M^{(K)}$ is a ``composite'' transfer matrix with the superscript $(K)$ denoting that it has been multiplied with the kernel matrix. $M_{\hat O_i}^{(K)} $ is the transfer matrix with an insertion of operator $\hat O_i$ in the physical legs. One may worry about potential sign problems since, for an odd number of bond modes, Eq.\ (\ref{eq:wavefunction_contraction}) has an extra parity matrix. In fact this does not influence the expectation values for the bonds inside the chain, and only possible signs will happen at the two ends of the chain if the fMPS chain is closed. In Eq.\,\eqref{eq:trans_mat_exp_values}, we denote the parity matrix of boundary bond modes as $\Xi$, and the exponent $N^{\vec t}$ is the half of the total number of bond modes in the MPS chain, or equivalently, the total occupation number of all kernel states used for contractions. The derivation of Eq.\eqref{eq:trans_mat_exp_values} is in Appendix \ref{appendix:transfer_mat_and_computation}.

\begin{figure}
    \centering
    \includegraphics[width=0.5\textwidth]{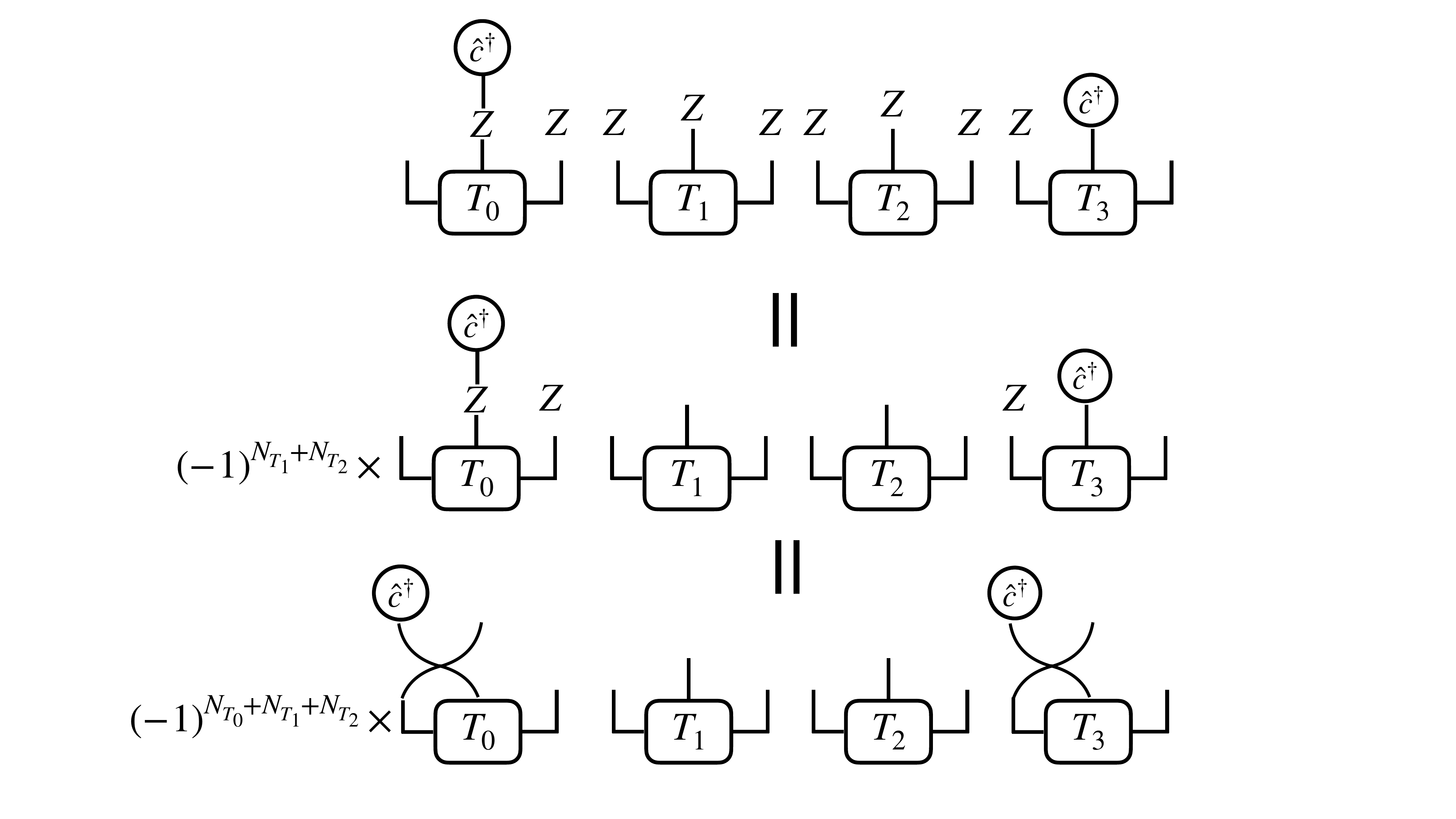}
    \caption{Simplifying the insertion of Jordan-Wigner string}
    \label{fig:JW_insertion}
\end{figure}

For the computation of non-local correlation functions, we need to insert suitable JW strings. For instance, to compute the two-point function $\ev{\hc^\dagger_i \hc_{i+n}}$, the usual way with one-dimensional MPS is to insert a JW string from the site $i$ to the site $i+n-1$. In our fMPS representation, since all local tensors are particle-number conserved fermion states, this JW string can be simplified. The expectation value is 
\begin{equation}
\begin{split}
        \ev{\hc^{\dagger}_i\hc_{i+n}} &= \Tr_{\text{phys}}\left[\hc_i^\dagger \hc_{i+n} \Tr_{\text{bond}}(\hat \rho_{\ket{\psi}}\hat \rho_{K})\right]\\
        &= \Tr_{\text{bond}}\Tr_{\text{phys}} \left[\hc^\dagger_i\hc_{i+n} \hat \rho_{\ket{\psi}}\hat \rho_{K}\right],
\end{split}
\end{equation}
where we use $\hat\rho_{\psi}$ to represent the density operator of the product state of all local tensors and $\hat\rho_{K}$ the product state of all contraction kernels. Then the JW string is going through all physical modes and bond modes from the mode $i$ to the mode $i+n$, as shown in Fig.\ \ref{fig:JW_insertion}. 
But all local tensors are particle-number conserved, so the action of JW-string, which is a string of local parity operators, on a local tensor simply returns an overall sign; see Fig.\ \ref{fig:JW_insertion}. 

\section{Construction of Gaussian fMPS}\label{sec:construct_fgmps}
Given a fermionic Gaussian state, we want to construct an MPS representation from the correlation matrix\cite{jin2020efficient_tu,fishman2015compression,PhysRevA.104.012415_Plenio}. As an illustration, consider a single band Hamiltonian of spinless fermions
\begin{equation}
    \begin{split}
            \hat H = -t\sum_{i} \hc^\dagger_{i} \hc_{i+1} + h.c. 
            = -\sum_{k} (2t\cos k) \hc^\dagger_{k} \hc_{k},
    \end{split}    
\end{equation}
which has the ground state
\begin{equation}
\ket{\psi} = \prod_{|k| \le \pi/2} \hc^\dagger_k \ket{0}.    
\end{equation}
This ground state is fully described by its correlation matrix 
\begin{equation}
    C_{i,j} = \bra{\psi}\hc_{i}\hc_j^\dagger \ket{\psi} = \Psi \Psi^{\dagger}
\end{equation}
where $\Psi$ contains the column eigenvectors of the single-particle Hamiltonian corresponding to all the filled modes.

We discuss two schemes for obtaining the MPS representation of the state. As usual, one can obtain an MPS representation by successively performing Schmdit decomposition along the chain~\cite{schuch_bauer_2019matrix,petrica2021finite,Jin_Hatree-Fock-Bogoliubov}
from, say, left to right. We will refer to this as ``Scheme I.'' Alternatively, we can also construct a translation-invariant MPS representation directly, which we refer to as ``Scheme II.'' As we will discuss in details below, Scheme I can faithfully capture the Fermi surface singularity of the original filled Fermi sea, but due to the lack of translation invariance our calculations are limited to relatively small system sizes. In contrast, Scheme II enables the evaluation of physical observables on much longer chains, but, due to the finite bond dimension, the resulting translation-invariant MPS has a smeared Fermi surface although the Fermi wave vector can still be readily recognized~\cite{PhysRevLett.96.010404_smoothed_FS,mortier2022tensor_resolve_FS}.

\subsection{Scheme I: Successive Schmidt Decomposition}\label{sec:fGMPS_scheme_I}
The first scheme is to successively perform the Schmidt decomposition discussed in Sec.\ \ref{sec:tool_fGMPS}. This can be summarized in four steps:
\begin{enumerate}
    \item Choose a subregion of length $L_c$. 
    Define a local Hilbert space $\mathcal H_x$ by combining all the existing bond modes with the physical modes in the sub region. 
    Diagonalize the corresponding reduced correlation matrix $(C)|_{i,j\in \mathcal H_x}$ with unitary matrix $U$. Denoting the eigenvalues as $\Lambda$, we have $(C)|_{i,j\in \mathcal H_x} = U\Lambda U^\dagger$.
    \item Freeze the eigenmodes in $\mathcal H_L$ with eigenvalues $\lambda< \ep$ or $\lambda>1-\ep$, where $\ep$ is a chosen entanglement threshold. 
    \item Use the kernel state to purify the unitary (or the isometry if some modes are frozen) 
    $U$ to get a Gaussian state $\ket{M_x}$. 
    \item Take the transformed correlation matrix as the new state with new bond modes defined from the Schdmit decomposition above. Repeat from step 1 until we have swept through the entire chain.
\end{enumerate}

\begin{widetext}
    \begin{figure*}
            \centering
            \includegraphics[width=0.9\textwidth]{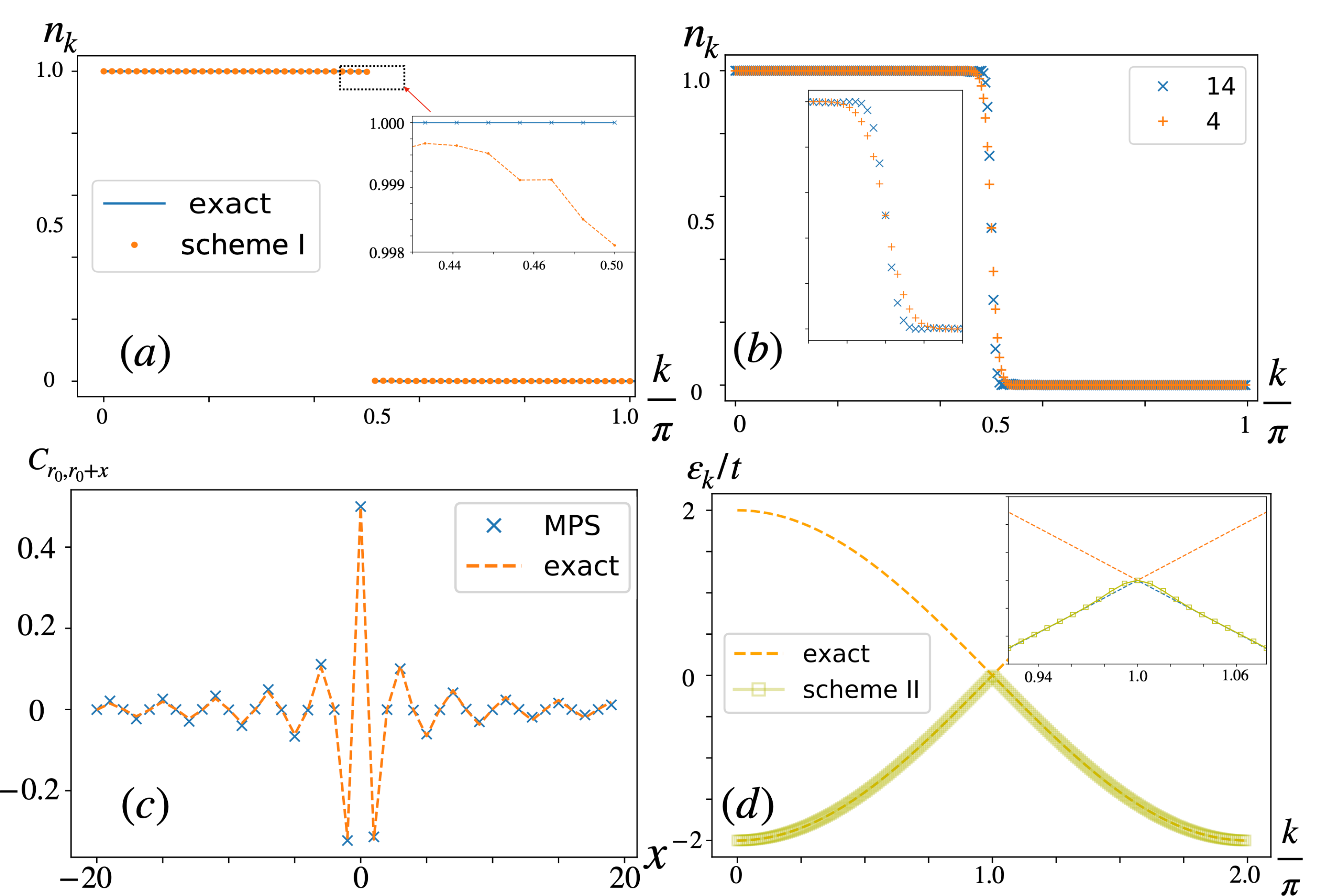}
            \caption{(a) The momentum distribution of the fermionic Gaussian matrix product state (fGMPS) constructed from Schmidt decomposition (Scheme I). The entanglement threshold is $10^{-3}$ and the chain is of length 128. (b) The momentum distribution of fGMPS constructed from tree-stack method. The number of bond modes is 14, and the length of the fGMPS is 500. (c) The crosses are the correlation functions of the translation-invariant fGMPS and the dashed lines are exact correlation functions. (d) The reconstructed dispersion for modes at each momentum from translation-invariant tensors (Scheme II), in the folded Brillouin zone. The zoomed in region shows the quadratic dispersion near the Fermi point.
            }
            \label{fig:gaussian_combined}
    \end{figure*}
\end{widetext}

By doing these steps successively, we can obtain a set of local fermionic Gaussian tensors $\{M_0, M_1, ..., M_\ell\}$, $\ell = L/L_c$. These local fermionic Gaussian tensors include left bonds modes, physical modes and right bond modes. Contracting all bond modes we can obtain the whole fGMPS state. Table \ref{tab:schemeI_summary} summarizes some numerical results on the MPS approximation of the exact state $\psi$. The momentum distribution $n_k = \ev{\hc_k^\dagger\hc}$ computed from the correlation matrix elements is shown in Fig.\ \ref{fig:gaussian_combined}(a).

\begin{center}
\begin{table}
\caption{Summary of performance in obtaining matrix product representation of a filled Fermi sea in Scheme I. The second column is the entanglement threshold for truncation, and the third column is the number of maximal bond modes.The fidelity is $|\ev{\psi_\text{fMPS}|\psi_{exact}}|$. The last two columns are energy densities (in units of $t$) computed from fMPS results and exact ground states. 
\label{tab:schemeI_summary}}
\begin{tabular}{|c|c|c|c|c|c|} 
 \hline
 $L_c$ & ent th& m.b.m & fidelity & $E_{M}/L$ & $E_{gs}/L$ \\ [0.5ex] 
 \hline\hline
1 & $10^{-2}$ & 4 & 0.580 & -0.6300 & \multirow{3}{*}{-0.6365} \\ 
 \cline{1-5}
2 & $10^{-2}$ & 4 & 0.662 & -0.6317 &  \\
 \cline{1-5}
1 & $10^{-3}$ & 7 & 0.953 & -0.6359 &  \\ 
 \cline{1-5}
2 & $10^{-3}$ & 7 & 0.974 & -0.6362 &  \\
 \hline
\end{tabular}
\end{table}
\end{center}

The advantage of this first scheme is that it is more straightforward, and is also more accurate in describing the low-energy behavior. For instance, the momentum distribution shows a sharp Fermi surface with a discontinuity. The shortcoming is that the obtained tensors are not translation-invariant, and it requires significantly more computational resources for large the system size in the subsequent many-body calculation after performing projection. Nevertheless, this construction provides some information about the correlation functions in a finite-size system, and can be regraded as a crosscheck before computing a larger system through Scheme II below.

\subsection{Scheme II: Translation-invariant tensors}
Alternatively, one could also construct translation-invariant MPS for the filled Fermi sea. This can be achieved by optimizing an MPS defined by local tensors with fixed, predefined bond dimensions to minimize the energy of a model Hamiltonian \cite{jin2020efficient_tu,mortier2022tensor_resolve_FS,li2023u_tu,yang2023projected_tu}.
Here, we adopt a more constructive approach in which the set of translation-invariant local tensors are obtained using the method introduced in Ref.\ \cite{he2024stacked}, which is dubbed ``tree-stack-compress.''
 
This approach returns a local tensor defined on a supercell of two physical sites, i.e., a local state defined in a Hilbert space of two physical modes together with some bond modes defined on both the left and right sides.
The resulting MPS state is then obtained by repeating this local tensor, and so it has discrete translation invariance in units of two sites. This subsequently saves computational cost when computing physical observables.

\begin{figure}[htpb]
    \centering
    \includegraphics[width=0.45\textwidth]{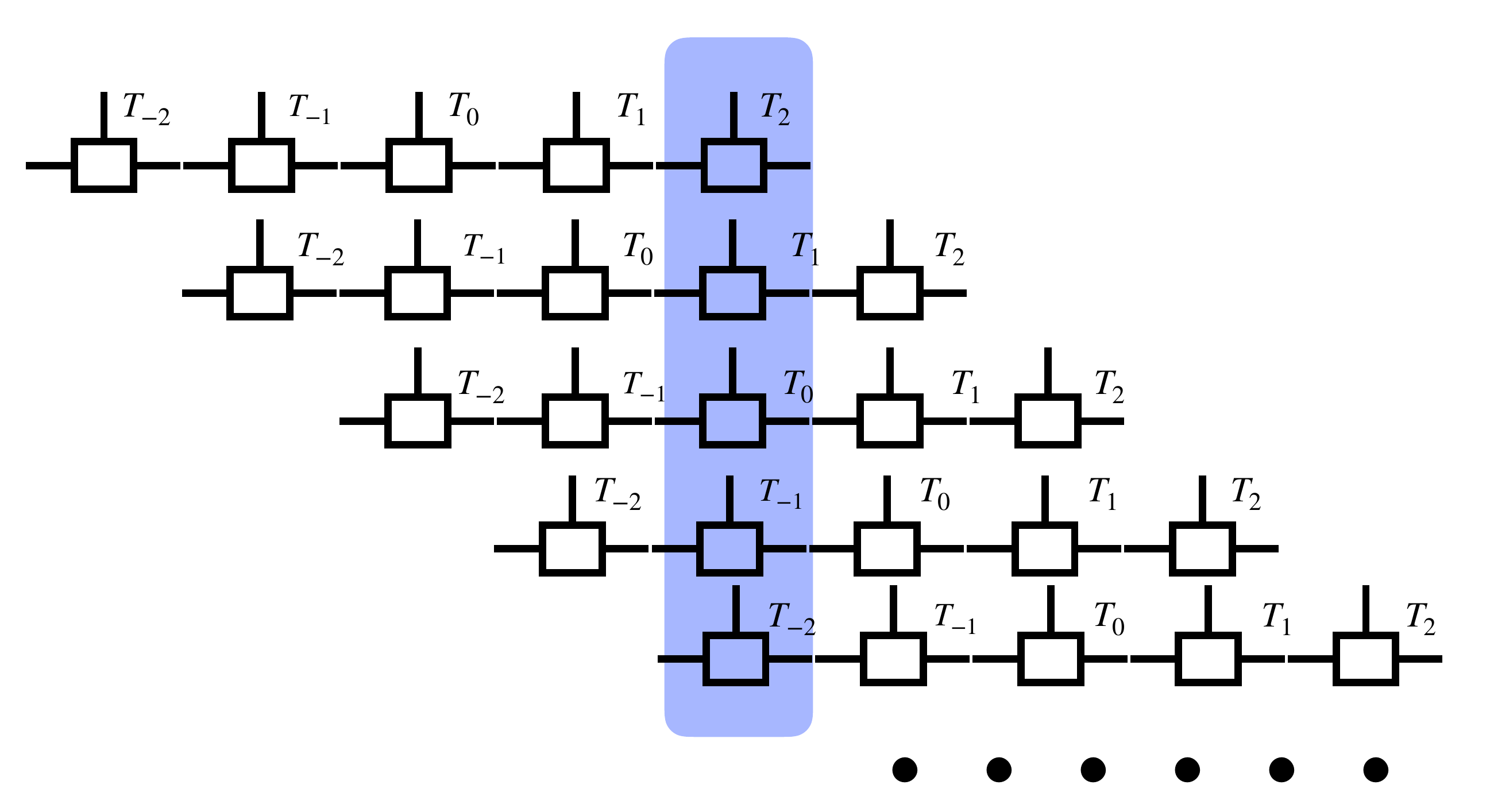}
    \caption{Schematic for the tree-stack construction of a free-fermion matrix product state. Each row represents a tensor network for a Wannier state with truncation radius 2. The colored region shows how the tensors are stacked to give a translation-invariant MPS approximation to the full state.}
    \label{fig:1d_stacking}
\end{figure}

We first treat the 1d spinless fermion chain using two-site unit cells. The original Brillouin zone is folded, and we get a two-band dispersion. The lower band is fully filled, which enables a Wannierization of the lower band. The tree-stack-compress method (Fig.\ \ref{fig:1d_stacking}) proceeds by first finding a fGMPS representation for each of the Wannier mode, and then stacking all their translation-related copies to obtain the full fGMPS. The bond dimension of the resulting fGMPS can then be brought down through a compression step.
Applied to the problem of a filled Fermi sea state, however, the Wannier modes are only algebraically instead of exponentially decaying. Nevertheless, one can still truncate the Wannier mode down to a finite region of some radius about the Wannier center, which amounts to approximating the filled Fermi sea by an atomic insulator. 
Such approximation works well for the high-energy (or short-distance) properties of the state, but is inherently incapable of capturing the Fermi surface singularity. 
Nevertheless, similar to existing works on the study of critical states using translation-invariant MPS~\cite{mortier2022tensor_resolve_FS,li2023u_tu}, a smeared Fermi surface is still retained and its sharpness is generically controlled by the bond dimension of the tensor-network representation. 
In our tree-stack-compress method, this is in turned controlled by the truncation radius used in obtaining the approximate Wannier functions. 

We summarize some numerical results here. For example, given the truncation radius of the Wannier function to be $16$, we can obtain a local tensor with $14$ bond modes on each side. The MPS representation is then further compressed such that each bond contains only four fermion modes.
For the compressed tensor, the difference of energy density from the exact ground state is $5\times 10^{-4}$. 
Its associated momentum distribution is shown in Fig.\ \ref{fig:gaussian_combined}(b), which shows a smeared Fermi surface discontinuity about $k_{\rm F} = \pi/2$. 
A comparison between the exact and MPS-approximated two-point functions is shown in Fig.\ \ref{fig:gaussian_combined}(c), which demonstrates good agreement in the short-distance.
The momentum-resolved energy spectrum for the fGMPS state is represented in Fig.\ \ref{fig:gaussian_combined}(d), which again shows good agreement with the exact result in the high-energy part of the spectrum, i.e., away from the Fermi surface. However, as shown in the inset, there is an inherent rounding of the linear dispersion about the Fermi surface, corresponding to its smearing. 

The Fermi surface smearing is a result of the critical nature of the filled Fermi sea state, which leads to a logarithmic contribution to the entanglement area law \cite{PhysRevLett.90.227902_vidal_2003,PhysRevLett.96.100503_entanglement_entropy_fermions_Gioev,PhysRevLett.96.010404_smoothed_FS}. 
As a result, the approximation of this critical state using an MPS with a finite bond dimension leads to an unavoidable failure in the long-wavelength limit. 
Nevertheless, the degree of smearing, i.e., the sharpness of the Fermi surface, can generally be controlled through the bond dimension \cite{geraedts2016half_zaletel_science,mortier2022tensor_resolve_FS}.
In Appendix~\ref{appendix:finite_bond_scaling}, we demonstrate how the smearing width can be systematically reduced at the cost of an increasing bond dimension.

\section{Gutzwiller projected wave functions}\label{sec:gutz_wave}

Having obtained the MPS approximation of the filled Fermi sea state, we can move on to construction of many-body-local tensors for interacting problems. One typical example is the Gutzwiller projected wave function~\cite{gutzwiller1963effect}. The standard Gutzwiller projector is 
\begin{equation}
    \hat G_{\rm double} = \prod_{\vec r} (1 - (1-g)\hat n_{\vec r,\uparrow},\hat n_{\vec r,\downarrow}),
\end{equation}
which for $g < 1$ penalizes configurations with double occupancy on any site. Here, we consider repulsive interactions between the fermions and so take $g\in [0, 1]$. 
The (unnormalized) Gutzwiller projected wave functions are defined as $\hat G\ket{\psi}$ where $\ket{\psi}$ is the filled Fermi sea state. In one spatial dimension, physical observables of the standard Gutzwiller projected states can be evaluated analytically for a finite $g$ when $\ket{\psi}$ is half-filled~\cite{gebhard1988correlation,metzner1988analytic, kollar2002exact}. This case then serves as a benchmark for our numerical methods. 

We would also consider another Gutzwiller-type projector, suitable for spinless fermion, which penalizes configurations based on nearest-neighbor (nnbr) repulsion:
\begin{equation}
\begin{split}
    \hat G_{\rm nnbr} &= \prod_{x} (1- (1-g)\hat n_x\hat n_{x+1} ).
\end{split}
\end{equation}
Notice that, as $\hat G_{\rm double}$ acts on different sites independently, its action on a MPS can be computed readily. In contrast, $\hat G_{\rm nnbr}$ imprints inter-site correlations on the state and does not act in a strictly local manner. Nevertheless, one can readily derive its matrix product representation, as is shown in Appendix \ref{appendix:joint_gutz}.
We note that earlier works have employed the Gutzwiller approximation to study this problem of spinless fermions projected by $\hat G_{\rm nnbr} $\cite{fazekas1972wigner,fazekas1989variational,seibold1997gutzwiller,PhysRevB.88.035114_spinless_gutz}. Here, we show that the tensor networks can provide an alternative numerical means for extracting physical observables like momentum distributions and static structure factors from the projected wave function.

\begin{figure}
    \centering
    \includegraphics[width=0.5\textwidth]{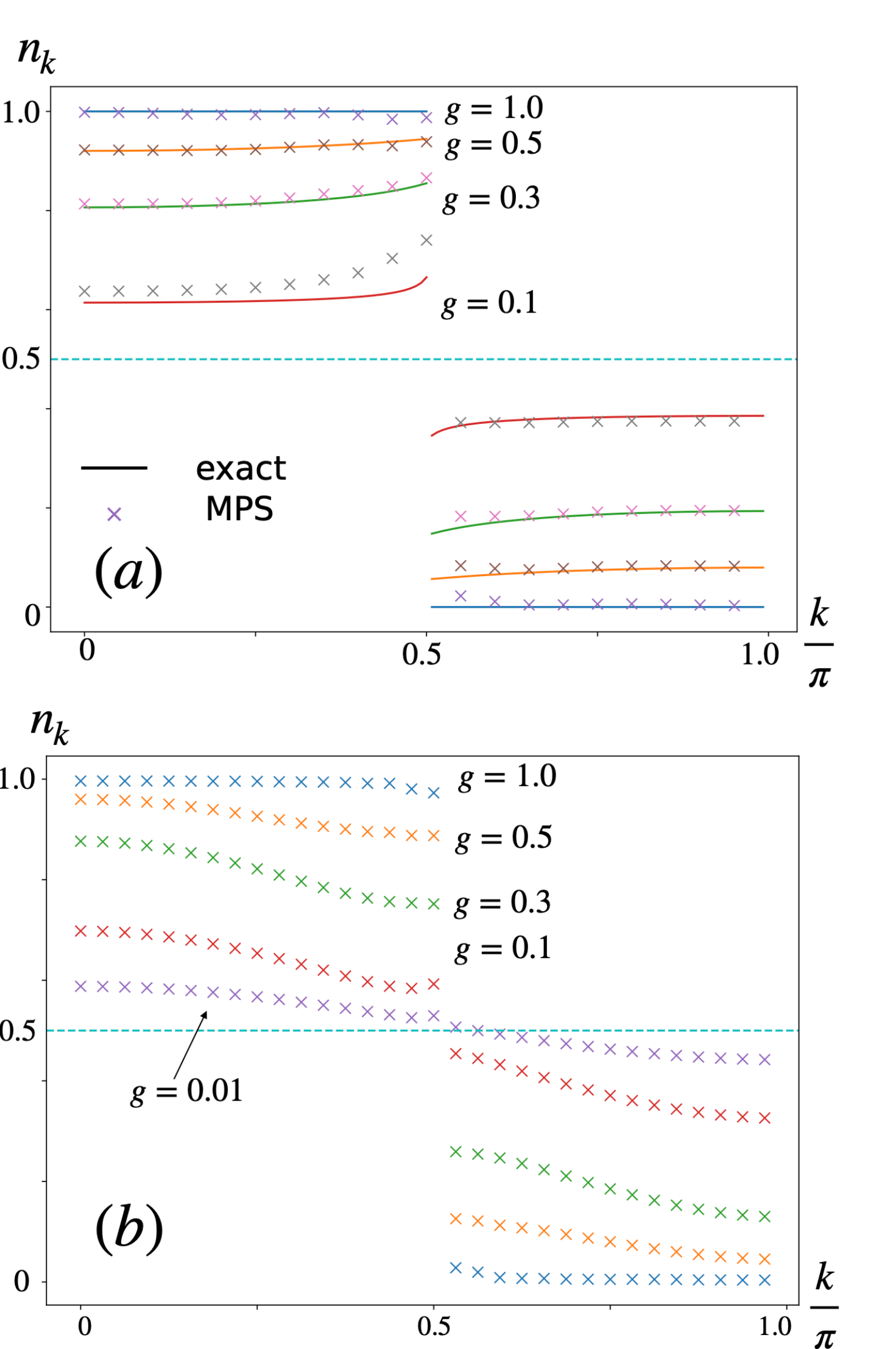}
    \caption{(a) The momentum distribution of the $\hat G_{\text{double}}$-projected state of 40 sites. The entanglement threshold of the fGMPS is $10^{-2}$ and the maximal number of bond modes in the chain is 3. (b) The momentum distribution for $\hat G_{\text{nnbr}}$-projected state in a chain of 64 sites}
    \label{fig:brute_force_nk}
\end{figure}

\subsection{Many-body MPS from Scheme I}
We begin by studying the projected Fermi sea states obtained by applying the Gutzwiller-type projectors to the fGMPS obtained through successive Schmidt decomposition (c.f.\ Sec.\ \ref{sec:fGMPS_scheme_I}). We compute various correlation functions for these projected states. 
First, we benchmark our results in the spinful Gutzwiller projector $\hat G_{\rm double}$ acting on the half-filled Fermi sea. 
The two-point functions $\ev{\hc^\dagger_{i,\uparrow}\hc_{j,\uparrow}} $ of a spinful fermion chain of $L=40$ sites for varying $g$ were computed. The momentum distribution $n_k$, obtained through Fourier transform of the two-point functions, is shown in Fig.\ \ref{fig:brute_force_nk}(a). 
In agreement with the exact results, we find that the Fermi surface discontinuity survives for finite $g$, although the size of the jump, corresponding to the quasi-particle weight, is reduced. In addition, the Gutzwiller projector modifies the momentum distribution far away from the Fermi surface; as we will see, such short-distance properties of the projected state can be well-reproduced using tensor-network methods.
Notice that, due to the exact degeneracy at the Fermi surface, we choose the occupation number per spin species to be $\frac{L}{2}+1$. In contrast, the analytical result is for exact half-filling. The difference in fermion count, which is a finite-size effect, leads to a more significant deviation between the computed and the exact analytical result as the projection strength increases (i.e., for smaller values of $g$). 

Likewise, in Fig.\ \ref{fig:brute_force_nk}(b) we present the momentum distribution of nearest-neighbor Gutzwiller projected state $\hat G_{\rm nnbr} \ket{\psi}$ of a $64$-site chain. Similar to the spinful case, the Fermi surface persists for finite values of $g$, with the jump size is again renormalized to a smaller value than one. In the hard-projection limit of $g=0$, the projected state is exactly a cat state $\frac{1}{\sqrt{2}}\left(\ket{0101...} \pm \ket{1010...} \right)$ if the system is exactly half-filled. Correspondingly, the momentum distribution will be exactly $\frac{1}{2}$ over the entire Brillouin zone. Our results are qualitatively similar to \cite{fazekas1989variational} where the author studied the $n_k$ by Gutzwiller approximation. We also included the case of exact half-filling in Fig.\ \ref{fig:spinful_combined}(f) for comparison with the subsequent calculations using Scheme II (see Sec. \ref{sec:many-body_computation}).

\subsection{Many-body MPS from Scheme II}\label{sec:many-body_computation} 
\begin{widetext}
    \begin{figure*}
    \centering
    \includegraphics[width=\textwidth]{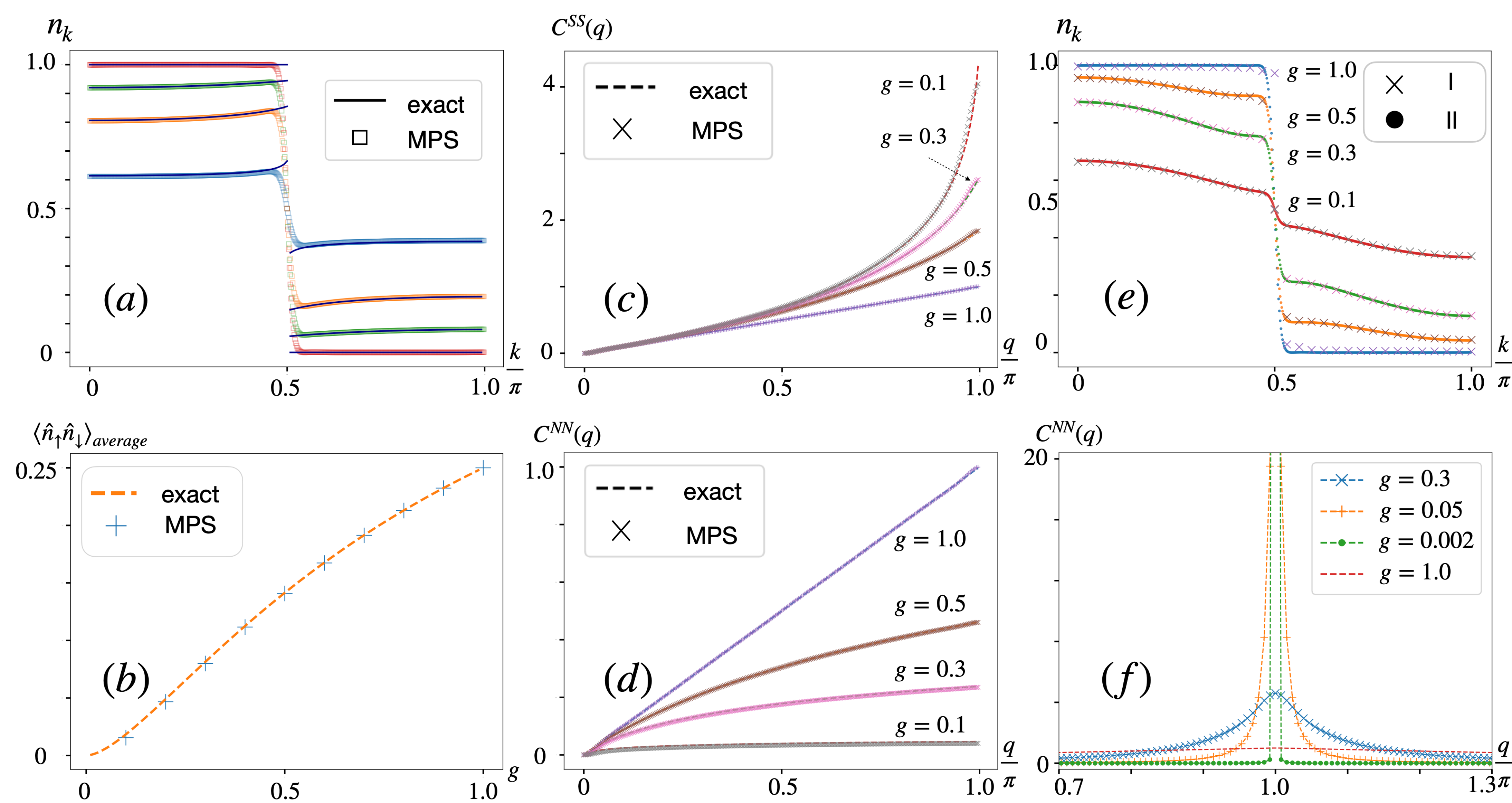}
    \caption{
    (a) The momentum distribution for spinful fermions with finite value of Gutzwiller projection $g$, based on spinless local tensor with $4$ bond modes on each side, and its comparison with the analytical results. The system size is $L = 1000$. (b) The average double occupancy compared with the analytical results. (c,d) The density-density correlation and spin-spin correlation in momentum space compared with the analytical results. (e) The momentum distribution of nearest-neighbor projected state. The crosses are computed with Scheme I (64 spinless fermions) and dots are computed from Scheme II (1000 spinless fermions). Both are exactly at half filling. (f) The density-density correlation function of nearest-neighbor projected state
    }
    \label{fig:spinful_combined}
\end{figure*}
\end{widetext}

While faithfully reproducing the Fermi surface discontinuity in the momentum distribution, the lack of translation invariance in the resulting tensor from the successive Schmidt decomposition (Scheme I) limits the system sizes we can reach. As an alternative, we now consider the projected Fermi sea state obtained from the translation-invariant local tensors described under Scheme II. 
First, we again benchmark our results against the standard Gutzwiller-projected state $\hat G_{\rm double}$, where we use a local tensor with $4$ bond modes on each side. The momentum distribution is shown in Fig.\ \ref{fig:spinful_combined}(a). We can compare it with the exact result as shown Fig.\ \ref{fig:spinful_combined}(d). We see they match well in the high-energy region far from the Fermi points. Near the $k_{\rm F} = \pi/2$, the fMPS results have smeared Fermi surfaces, while the exact results has a sharp Fermi surface. 

Another natural quantity to consider for the Gutzwiller projected wave function is the average double occupancy $\ev{\hat n_\uparrow\hat n_\downarrow}_{average} = \frac{1}{L}\sum_{l=1}^L\langle \hat{n}_{l,\uparrow}\hat{n}_{l,\downarrow}\rangle$, which is suppressed from the original filled Fermi sea.
Fig.\ \ref{fig:spinful_combined}(c) shows the exact average double occupancy in \cite{metzner1988analytic} compared with our numerical results for $g$ from $0.1$ to $1$ with step size of $0.1$, which accurately approached the exact results.
Other physical correlation functions, such as spin-spin correlation and density-density correlation,
\begin{equation}
\begin{split}
        C^{SS}(r) &= \frac{1}{L}\sum_{i}(\ev{\hat S^z_{i}\hat S^z_{i+r} } - \ev{\hat S^z_i}\ev{\hat S^z_{i+r}}) \\
        C^{NN}(r) &= \frac{1}{L}\sum_{i}(\ev{\hat n_{i} \hat n_{i+r} } - \ev{\hat n_i}\ev{\hat n_{i+r}}) \\
\end{split}
\end{equation}
can also be computed. After Fourier transformation, we obtain their representation in momentum space $C^{SS}(q)$ and $ C^{NN}(q)$, known as the static structure factor for spin and charge (see, for example, \cite{mahan}).
 In Fig.\ \ref{fig:spinful_combined}(b), the fMPS and exact spin-spin correlation in momentum space for $g = 0.1, 0.3, 0.5, 1$ are computed. When the momentum $|k|<\pi = 2 k_{\rm F}$, the fMPS results accurately fit the exact results with all computed $g$. When $|k|$ approaches $\pi$, the exact spin-spin correlation for $g\neq 1$ becomes a kink, which is the $2k_{\rm F}$ singularity. The fMPS results quantitatively are still close to the exact data but the cusps at $q = \pi$ are rounded. Fig.\ \ref{fig:spinful_combined}(e) is the comparison between fMPS and exact density-density correlations. Unlike the case for spin-spin correlations, when $g\neq 1$ the fMPS results accurately resolve the density-density correlations for the whole momentum space.

After benchmarking against the exact analytic results in the standard one-dimensional Gutzwiller projected Fermi sea, we move on to consider other projected Fermi sea states. The momentum distribution of the nearest-neighbor Gutzwiller projector $\hat G_{\rm nnbr}\ket{\psi}$ for spinless fermions is presented in 
Fig.\ \ref{fig:spinful_combined}. Up to the mentioned smearing of the Fermi surface singularity, the results are quantitatively consistent with the computations done with Scheme I, except that a much longer chain can now be assessed.
The density-density correlation function is computed for strong projections, which shows a peak at $2k_{\rm F} = \pi$, which indicates the symmetry breaking at $g = 0$; see Fig.\ \ref{fig:spinful_combined}(f). 

\begin{figure}
\includegraphics[width=0.4\textwidth]{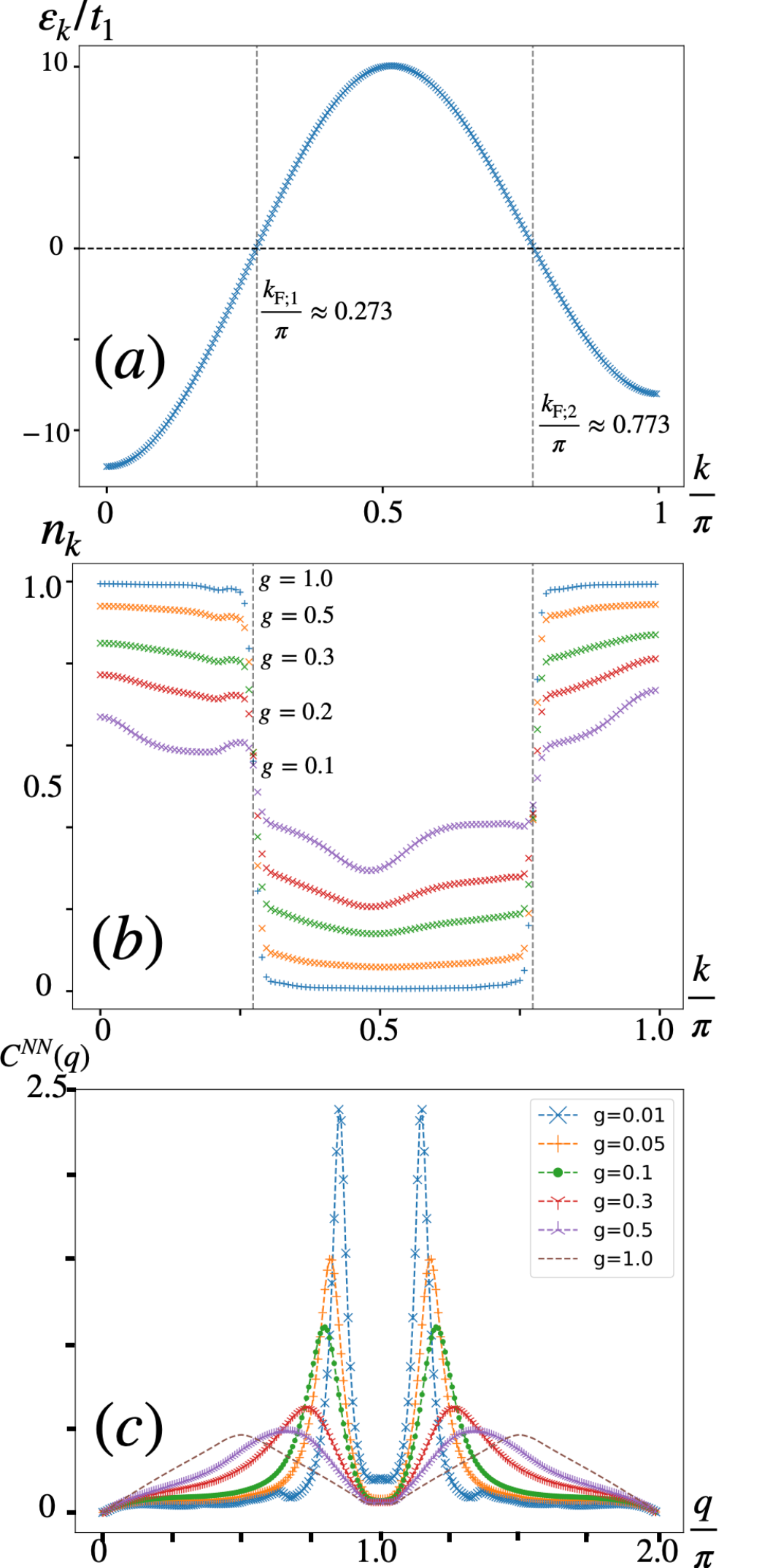}
\caption{
(a) The dispersion relation of Eq.\ \eqref{eq:ham_dfs} for $t_2 = 5 t_1$. (b) The momentum distribution of $\hat G_{\rm nnbr}$ projected wave function of 240 spinless fermions. (c) The density-density correlation function for $\hat G_{\rm nnbr}$ projected spinless fermions with two Fermi surfaces, for larger $g$ and smaller values.
}
\label{fig:dfs}
\end{figure}

A similar but more interesting problem is the projected wave function of a ground state with two pairs of distinct Fermi points. As an example, we take the $\ket{\psi}$ to be the ground state of the following next-nearest-neighbour hopping Hamiltonian,
\begin{equation}\label{eq:ham_dfs}
    \hat H = -\sum_{i=1}^L ( t_1 \hc^\dagger_i\hc_{i+1} + t_2\hc^\dagger_{i}\hc_{i+2} + h.c. )
\end{equation}
with $t_2 = 5t_1$. 
The solutions and correlation functions of this model are summarized in Appendix~\ref{appendix:t1_t2_model}.
The dispersion relation of this Hamiltonian is $\ep_k = -(t_1 \cos k + t_2\cos2k)$, as shown in Fig.\ \ref{fig:dfs}(a). By adjusting the chemical potential we get a ground state exactly at half-filling. For the chosen parameters here,  the two distinct Fermi points in $[0,\pi]$ are $k_{\rm F; 1}\approx 0.273\pi, k_{\rm F;2}\approx 0.773\pi$. We use the method of Scheme II to construct the translation-invariant local tensor, which has four physical modes and four bond modes for each side. With these Gaussian tensors, we compute the correlations for $\hat G_{\rm nnbr}\ket{\psi}$.
In Fig.\ \ref{fig:dfs}(b) the momentum distribution is shown for the projected wave functions with different $g$. Similar to the cases of a single Fermi sea, these projected states with two Fermi seas also maintains smeared Fermi surface discontinuities. 

We also study the density-density correlation function. For different values of $t_2/t_1$, the free fermion state before projection shows different features, see Fig.\ \ref{fig:dfs_folding_gaussian_nnq}(b) in Appendix \ref{appendix:t1_t2_model}. 
For different $g$, the computation of density-density correlation functions is shown in Fig.\ \ref{fig:dfs}(c). At the Gaussian limit ($g=1$), the two peaks are at $\pi/2$. As $g$ decreases, corresponding to increasing projection and hence correlation strength, the two peaks sharpen and develop in magnitude. Their positions also shift towards $q=\pi$. 
Such correlation-driven enhancement in the static structure factor can be interpreted as a precursor for a charge-density wave instability. Interestingly, we find that the characteristic wavelength of this charge density modulation is continuously renormalized by the correlation strength, although the location of the Fermi points, as shown in Fig.\ \ref{fig:dfs}(b), remains unchanged from the parent free-fermion state.

The behavior of the system in the limit of vanishing $g$ warrants further discussion.
In fact, when $g=0$ the exact projected state vanishes, as the input free-fermion state does not have any weights in the two states $\ket{1010...}$ and $\ket{0101...}$ (Appendix \ref{appendix:vanishing_norm}). However, in our fGMPS representation of the input state, small weights in these two components are generated due to the approximations involved. As $g$ vanishes, the relative weights of these unphysical components are generally amplified, which renders these results unreliable. For completeness, we show the behavior of $C^{NN}(q)$ for $g < 0.01$ in Appendix \ref{appendix:vanishing_norm}, but for the mentioned reasons these results in the $g \rightarrow 0$ limit should be interpreted with caution.

\section{Discussion}
In this work, we formulate a constructive approach for evaluating physical observables in one-dimensional projected Fermi sea states using matrix product states. As a benchmark, we show that our results are in good agreement with the exact analytical results available for half-filled spin-$1/2$ electrons subjected to the Gutzwiller projector which penalizes double occupancy. Notably, when translation-invariant states are used, the Fermi surface singularity is generally smeared out, as expected due to the short-range entangled nature of the matrix product ansatz. Nevertheless, the remnant of the Fermi surface is still discernible in physical observables like momentum distribution function and spin-spin or density-density correlation functions. We further apply our approach to study the projected Fermi sea states of spinless fermions at half filling and subjected to nearest-neighbor repulsion. For the case of a single pair of Fermi points, the density-density correlation shows a peak at $q=\pi$ as the projection becomes stronger, whereas for the case of two pairs of Fermi points, the two peaks originating from the pre-projection parent state develop in magnitude and shift towards 
$q=\pi$.

While we focus on the evaluation of physical observables in a pre-defined projected Fermi sea state in this work, our results can serve as the starting point for further variational calculations. 
Conceptually, the Gutzwiller-type projectors we considered can be readily generalized to any other short-range entangled operators admitting a matrix product representation. The projected states we constructed, which retain smeared Fermi surfaces, can be taken as the initial state of well-established variational energy minimizers, like the density-matrix renormalization group approach \cite{jin2021density_tu, petrica2021finite,Jin_Hatree-Fock-Bogoliubov}. 
We note that in TN methods like the density-matrix renormalization group, the energy of a Hamiltonian is minimized over the variational TN ansatz. For one-dimensional fermions subjected to local interactions, one generically expects the non-interacting ground state, with a sharp Fermi surface, to be superseded by a Luttinger liquid with only a kink singularity instead of a discontinuity in the the momentum distribution function. In contrast, our approach is constructive in nature and corresponds to a TN representation of the projected Fermi sea states. As a result, the Fermi surface discontinuity is approximately retained, in the sense that a sharp transition of momentum occupation at the Fermi surface, with the sharpness controlled by the bond dimension of the TN, is reproduced in our method.

We also note that our approach can be applied to study gapless phases of fermions in higher dimensions, possibly in the context of a projected parton state \cite{liu2012gutzwiller_tu,wu2020tensor_mpo_mps_tu,jin2021density_tu,petrica2021finite,Jin_Hatree-Fock-Bogoliubov,budaraju2024simulating}. 
For critical states with a vanishing Luttinger volume, our approach~\cite{he2024stacked}
for constructing a tensor-network representation of a gapless Gaussian state proceeds similarly in high dimensions; for more general states with a Fermi sea, however, our current approach is limited to commensurate cases for which the Luttinger volume vanishes under suitable Brillouin zone folding. Nevertheless, it is an interesting question to study the efficacy of the outlined approach in describing highly entangled quantum phases of matter, say topologically ordered system with spinon Fermi surfaces. Along another line, one may also combine the approach with entanglement renormalization
\cite{PhysRevLett.99.220405_MERA,PhysRevB.78.205116_gu_TERG,qi2013exact,wong2022zipper}, which can provide a more faithful tensor-network representation of gapless states with sharp Fermi surfaces.\\

\noindent {\itshape Note added.} 
In finalizing the manuscript, we became aware of the work \cite{tu_2408}, which also discussed how to construct fermionic Gaussian MPS from correlation matrices with mode truncation.

\begin{acknowledgements}
    This work is supported by the Hong Kong Research Grant Council (ECS 26308021) and the Croucher Foundation (CIA23SC01).  We thank Hong-Hao Tu for helpful discussions and comments.
\end{acknowledgements}

\appendix
\section{Exact solutions of 1d Gutzwiller projected state}
In the main part, we compared the numerical results with exact solutions for different correlation functions. Here we list the related formulas of the exact solutions based on \cite{gebhard1988correlation,metzner1988analytic,kollar2002exact}. 
The exact momentum distribution formula used in Fig.\ \ref{fig:spinful_combined}(d) is:
\begin{equation}
\begin{aligned}
    n_k&(g) = \frac{g^2+4g+1}{2(1+g)^2} + \frac{g^2}{(1+g^2)^2} \times \\
    \times &\begin{gathered}\frac{4}{\pi \sqrt{(2-G)^2-(\Tilde{k}G)^2}} \mathcal{K}(\frac{G\sqrt{1-\Tilde{k}^2}}{\sqrt{(2-G)^2-(\Tilde{k}G)^2}})
\end{gathered}
\end{aligned}
\end{equation}
where 
$\Tilde{k} = 2|k|/\pi\leq 2$, $G = 1-g^2$ and 
\begin{equation}
    \mathcal{K}(k) = \int_0^{2\pi}d \psi[1-k^2\sin^2(\psi)]^{-1/2}
\end{equation}
is the first kind of complete elliptic integral.\par
The following is the formula for the spin-spin correlation function in real space $C^{SS}(r) = \frac{1}{L}\sum^{L}_{l=1}(\langle\hat{S}_{r+l}^z\hat{S}_{l}^z\rangle_G-\langle\hat{S}_{r+l}^z\rangle_G\langle\hat{S}_{l}^z\rangle_G)$
for electron density $n= n_\uparrow + n_\downarrow=1$:

\begin{equation}
    C^{SS}_{j\geq 1}( n = 1) =-\frac{1}{\pi j}\int_0^1 dy\frac{\sin(\pi jy)}{F(y)}
\end{equation}
where the function $F(x) \equiv 1-(1-g^2)x$. The formula comes from the Fourier transform of the spin-spin correlation in momentum space: 
\begin{equation}
    C^{SS}(k) = 
    \begin{cases}
        -\frac{1}{1-g^2}\ln{F}(\frac{|k|}{\pi}), 0 \leq|k|\leq2k_{\rm F}\\
        -\frac{1}{1-g^2}\ln{F}(n), 2k_{\rm F}\leq|k|\leq \pi
    \end{cases}
\end{equation}
The density-density correlation in real space is defined as $C^{NN}(r) = \frac{1}{L}\sum^{L}_{l=1}(\langle\hat{n}_{r+l}^z\hat{n}_{l}^z\rangle_G-\langle\hat{n}_{r+l}^z\rangle_G\langle\hat{n}_{l}^z\rangle_G)$. Upon Fourier transform to the momentum space, one finds
\begin{equation}
    C^{NN}(k;g,n=1) = \frac{g^2}{1-g^2}\ln{F}(-\frac{Q}{g^2})
\end{equation}
the exact correlation in real space is:
\begin{equation}
\begin{aligned}
    C_j^{NN} = &
    -\frac{\Bar{p}}{(\pi j)^2}\{\sin{(\pi j) }\ln{g^2} + \sin{\bar{p}}[Ci(\bar{p})-Ci(p)]+\\
    &+\cos{\bar{p}}[Si(p)-Si(\bar{p})]\}
\end{aligned}
\end{equation}
where $p = \pi j /(1-g^2)$, $\bar{p} = g^2 p = \pi j g^2/(1-g^2)$, $Si(x) = \int_0^{x}\frac{\sin t}{t}d t$ is the sine integral function and $C_i(x) = -\int_x^infty \frac{\cos{t}}{t}dt$ is the cosine integral function.\\
The formula for double occupancy formula in real space is:
\begin{align}
    &C^{D} = \frac{1}{L}\sum_{l=1}^L\langle \hat{n}_{l,\uparrow}\hat{n}_{l,\downarrow}\rangle_G =\\ &\frac{g^2}{2(1-g^2)^2}[-(1-g^2)(n-m)+\ln{\frac{1-(1-g^2)m}{1-(1-g^2)n}}]
\end{align}
where $m = (n_\uparrow-n_\downarrow)$ is the magnetization and $n$  is the electron density.

\section{$\hat G_{\rm nnbr }$-Projected state at $g=0$  }\label{appendix:vanishing_norm}

The nearest-neighbor Gutzwiller projector $\hat G_{\rm nnbr}$ tends to suppress the weight of configurations with doubly occupied nearest-neighbor sites in the state. At the limit $g=0$, only configurations like $\ket{0101...}$ and $\ket{1010...}$ will survive. Such single configuration has exponentially small overlap with the pre-projected state, which we can show by a simple estimation as follows. Given the fermionic Gaussian state $\ket{\psi} = \prod_k{\hc_k^\dagger}\ket{0}$ is at half-filling, the component $\ket{0101...}$ will have a coefficient determined by the Slater determinant
\begin{equation}
 \ev{0101... | \psi} = \frac{1}{L^{L/4}}\sum_{\sigma\in \mathcal{S}_{L/2} } {\rm sgn}(\sigma) e^{i\sum_j x_j \sigma(k_j)},
\end{equation}
where $\sigma\in \mathcal{S}_{L/2}$ denotes the permutation of $L/2$ objects, and $\sigma(k_j)$ denotes the permutation of all momentum indices of modes in the Fermi sea. Because the norm of each summand is not greater than 1, we see
\begin{equation}
    |\ev{0101... | \psi}| \leq \frac{2^L}{L^{L/4}} = e^{L(\text{ln}2 - \frac{1}{4}\ln L)} < e^{-\chi_0 L},
\end{equation}
as long as $L$ is large enough there always exist a positive constant $\chi_0$ satisfying the inequality. 
For the case with only nearest-neighbor hopping, there is only one pair of Fermi points, which does contain the components $\ket{0101...}$ and $\ket{1010...}$. But for the case with next-nearest-neighbor hopping the thing is different by the following claim.

{\bfseries Claim}: The overlap of $\ket{1010...}$ and the two-pair-Fermi-point state $\ket{\psi}$ as the ground state of the next-nearest-hopping model is zero.\\

{\bfseries Proof}: First we notice the eigenmodes of the two-pair-Fermi-point state are $\hc^\dagger_k = \sum_{k}e^{ikx}\hc^\dagger_x$. The coefficient $\langle 1010...\ket{\psi}$ is computed from the Slater determinant of the submatrix containing $x=0,2,4,...$. Restricting the coefficients of $\hc^\dagger_k$ to these sites, we see they form a vector 
\begin{equation}
    (1, e^{2ik}, e^{4ik}, ... )^T.
\end{equation}
This vector is linear dependent to another vector with $k'=k+\pi$. So if there are at least one pair of such $k,k'$ in the Fermi sea, then the determinant will be exactly zero.
Now we notice an important fact: the next-nearest hopping model has always a pair of such modes in the two disconnected pieces of Fermi sea, $(k,k')=(0,\pi)$, whatever $t_1/t_2$ is chosen as long as there are two pairs of Fermi points. Similarly for the overlap $\langle 0101...\ket{\psi}$.

\begin{figure}
    \centering
    \includegraphics[width=0.5\textwidth]{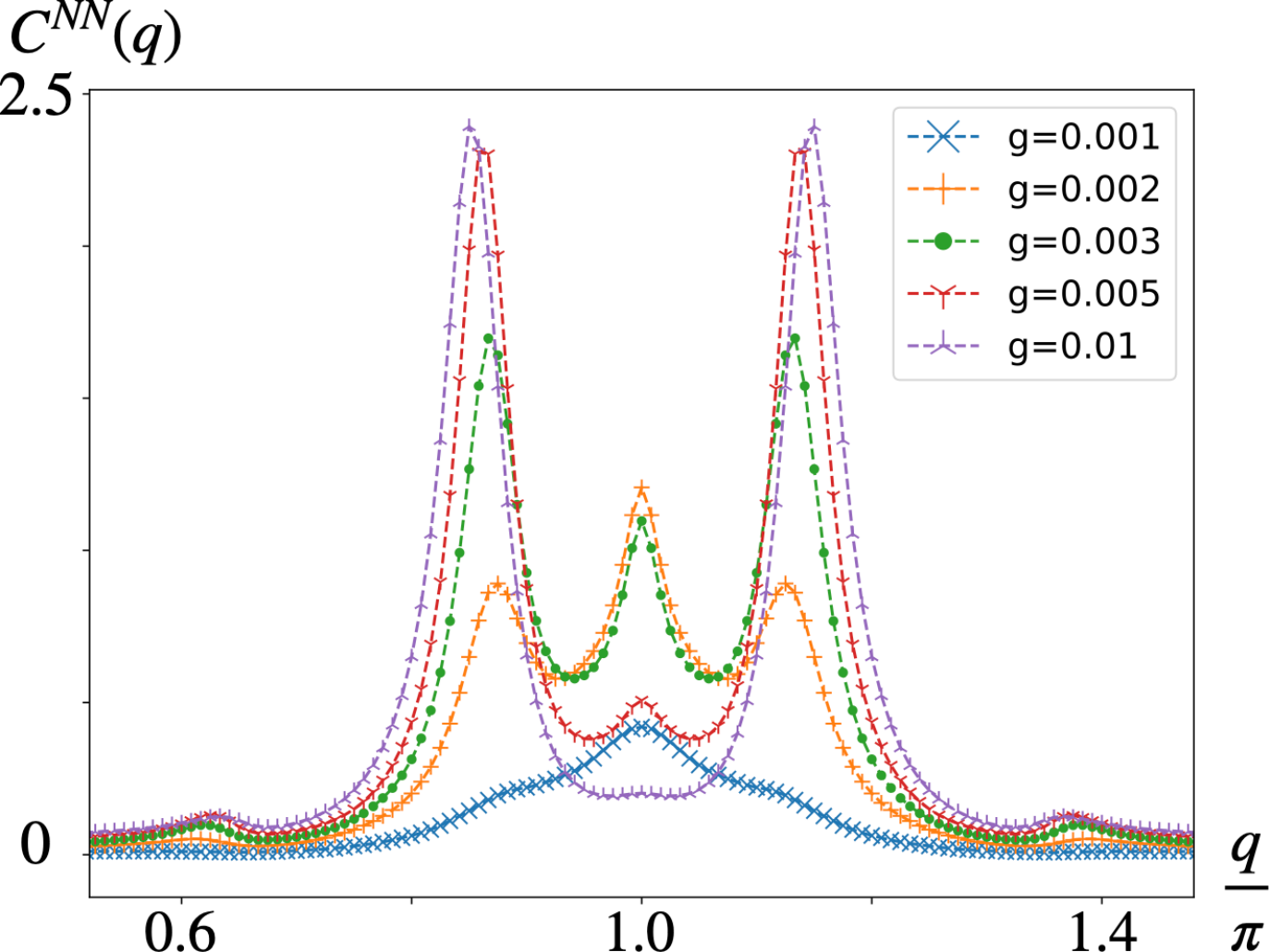}
    \caption{The static structure factor for $g<0.01$ as a zoomed-in style.}
    \label{fig:cnn_small_g}
\end{figure}

One the other hand, our Wannierized wave functions might contain some component of such $\ket{1010...}$ states. These are artificial components introduced in the construction of fGMPS. As discussed in the main text, the projected state for $g<0.01$ may have amplified unphysical components. The plot of the $C^{NN}(q)$ is shown in Fig.\ \ref{fig:cnn_small_g}.

\section{The next-nearest-neighbor hopping model}\label{appendix:t1_t2_model}
In momentum space, the model in Eq.\ \eqref{eq:ham_dfs} reads
\begin{equation}\label{eq:dfs_band_structure}\begin{split}
    \hc^\dagger_k =& \frac{1}{\sqrt{L}}\sum_{i} e^{ikx_i} \hc_i^\dagger, \\
    \hat H =& -\sum_{k} (t_1\cos k +t_2\cos2k) \hc_k^\dagger \hc_k.
\end{split} \end{equation}
Our motivation for introducing the next-nearest-neighbor hopping is to obtain a filled Fermi sea state with multiple pairs of Fermi points. To this end, we can first solve for the critical point for which the local band minimum at $k = \pi$ becomes degenerate in energy with the original Fermi level at $k=\pi/2$ at half-filling 
\begin{equation}
    t_1\cos\frac{\pi}{2} + t_2\cos\pi = t_1\cos \pi + t_2\cos2\pi \Rightarrow t_1 = 2t_2.
\end{equation}
This means we need $t_1<2t_2$ to obtain a new pair of Fermi points. Denote $t_1/t_2 = t < 2$, and assume the chemical potential is $\mu$, we can solve the following equation to determine the Fermi points:
\begin{equation}
    t_1\cos k+ t_2(2\cos^2 k-1)  +\mu = 0
\end{equation}
which leads to 
\begin{equation}
    \cos k_{1,2} = \frac{-t\pm\sqrt{t^2+8(1-\mu_t)}}{4}, \quad \mu_t = \mu/t_2.
\end{equation}
Suppose the roots of this equation are $0<k_{\rm F;1}<k_{\rm F;2}$, the condition for having two pairs of Fermi points is 
\begin{equation}
    k_{\rm F;2}-k_{\rm F;1} = \frac{\pi}{2} \Longleftrightarrow 
    \cos k_{\rm F;1}\cos k_{\rm F;2} = -\sin k_{\rm F;1}\sin k_{\rm F;2},
\end{equation}
which can be simplified into 
\begin{equation}
   \mu_t = \frac{t^2}{4}.
\end{equation}
Then we know the solution should be 
\begin{equation}
    \cos k_{1,2} = \frac{-t\pm \sqrt{8-t^2} }{4}.
\end{equation}
The condition of having two different roots is 
\begin{equation}
    \cos k_{\rm F;2}\le 0\le \cos k_{\rm F;1},  \mu_t \geq t-1,
\end{equation}
the latter is coming from the $-1\le \cos k\le 1$, so $t\le 2$ is obtained, consistent with the computation of critical point.
The Fermi seas are consisting of modes of $|k|<k_{\rm F;1}$ and $ k_{\rm F;2}<|k|<\pi $, so the ground state is $\ket{\psi} =\prod_{k\in FS} \hc^\dagger_k\ket{0} $. The correlation matrix in real space can be computed through the Wick's theorem. The result is 
\begin{equation}
    \ev{\hc_{i}\hc_{i+z}^\dagger } = \frac{1}{2}\delta_{z=0}-\delta_{z\neq 0} \frac{1}{L} \sum_{k\in DFS} e^{ikz}.
\end{equation}
By Wick's theorem, the density-density correlation function can be computed. 

\begin{figure}
    \includegraphics[width=0.5\textwidth]{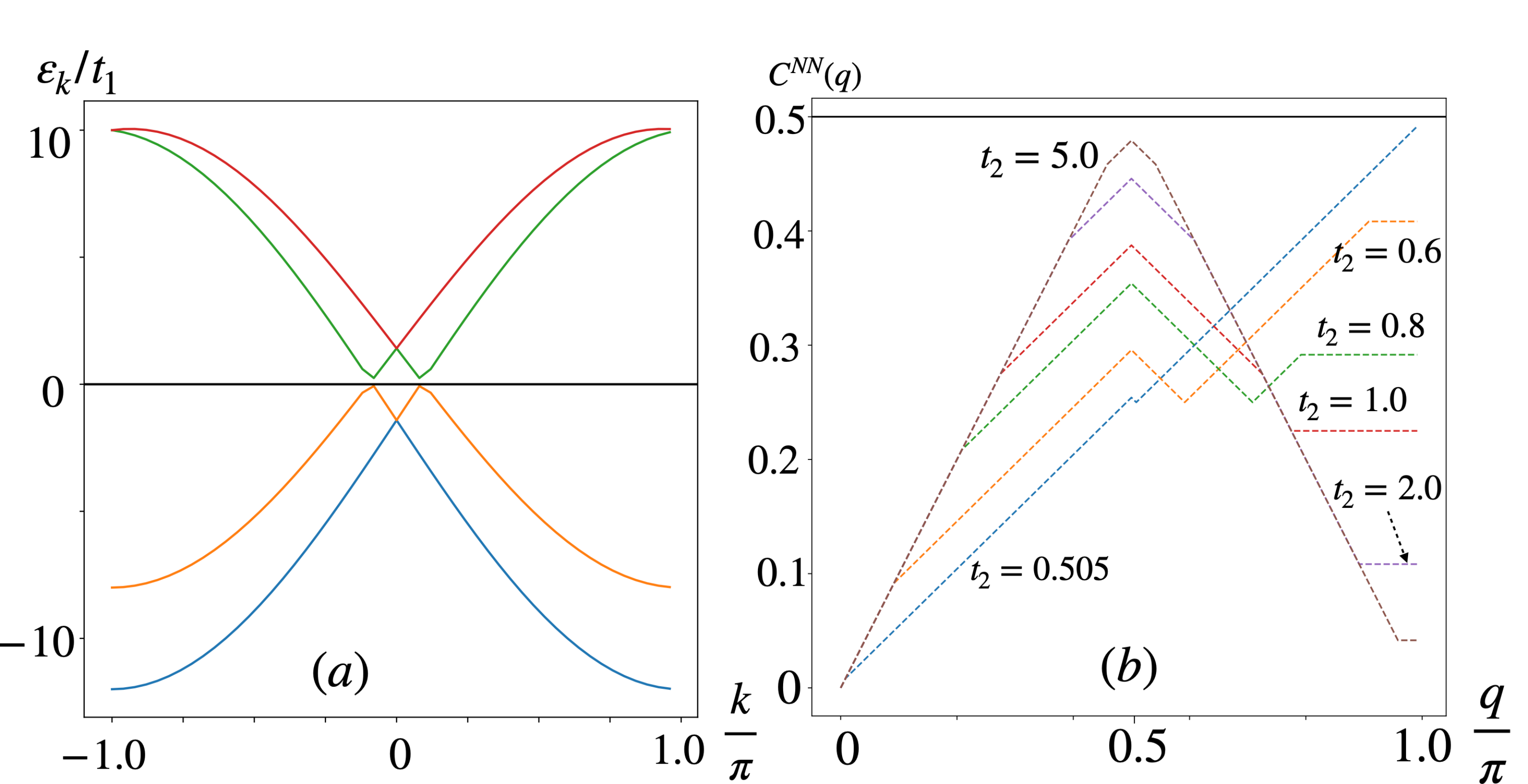}
\caption{(a) The band structure Eq.\ \eqref{eq:dfs_band_structure} in folded Brillouin zone for Wannierization in tree-stack construction of fGMPS; (b) The density-density correlation $C^{NN}(q)$ for different $t_2$ with fixed $t_1=1$. }
\label{fig:dfs_folding_gaussian_nnq}
\end{figure}

To study this in tree-stack method\cite{he2024stacked}, we need to first fold the Brillouin zone twice so that all filled modes consist of two fully filled bands (Fig.\ \ref{fig:dfs_folding_gaussian_nnq}(a)). Then do Wannierization and apply tree-stack construction, we can obtain a local tensor with four spinless fermions.

\section{Nearest-neighbor Gutzwiller Projector}\label{appendix:joint_gutz}
\begin{figure}[htpb]
    \centering
    \includegraphics[width=0.45\textwidth]{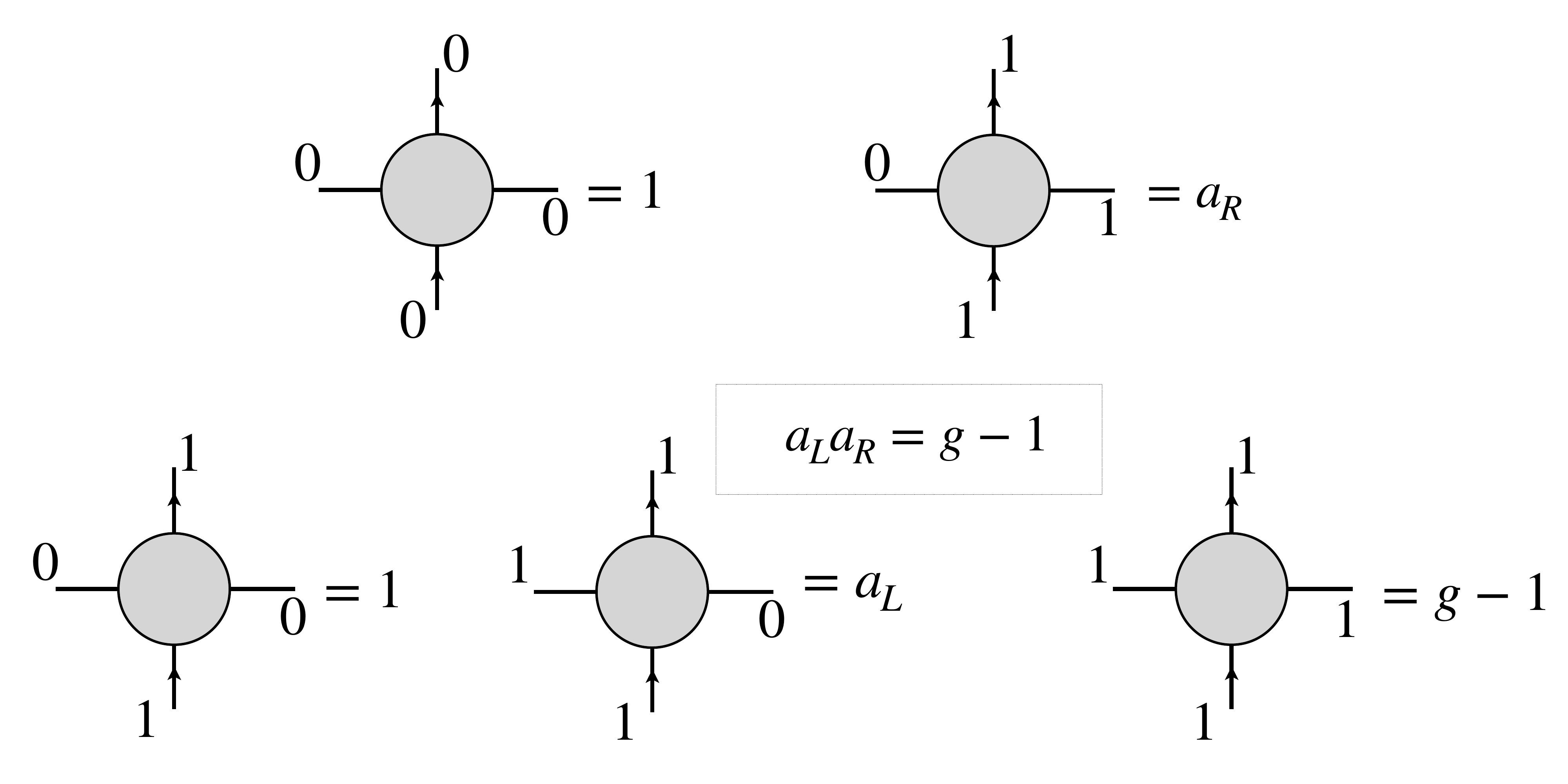}
    \caption{ Matrix product form for a Gutzwiller-like projector corresponding to nearest-neighbor density-density interaction. The vertical legs correspond to the physical fermionic Fock space (up = ``out" and down = ``in"). and the horizontal legs are virtual bosonic Hilbert space (of a qubit). All the configurations not shown have $0$ weight.}
    \label{fig:joint_gutzwiller_projector}
\end{figure}

Besides the Gutzwiller projected ground state of a spinful nearest-neighbor interaction chain, we also considered a free Fermi state 
projected by a nearest-neighbor Gutzwiller-like projector:
\begin{equation}
    \hat{G}_{\text{nnbr}} = \prod_xe^{-\alpha\hat{n}_x\hat{n}_{x+1}},
\end{equation}
where we could impose periodic boundary conditions. To unpack, we may focus on a single term in the product, and notice
\begin{equation}
    e^{-\alpha \hat{n}_x\hat{n}_{x+1}}\mapsto
    \begin{cases}
        1 &\text{for}(n_x,n_{x+1}) = (0,0),(0,1),(1,0);\\
        e^{-\alpha}&\text{for}(n_x,n_{x+1}) = (1,1)
    \end{cases}
\end{equation}
where we denote $g := e^{-\alpha}$ for convenience, and we may write
\begin{equation}
    e^{-\alpha \hat{n}_x\hat{n}_{x+1}} = 1 + (g-1)\hat{n}_x\hat{n}_{x+1}.
\end{equation}
This Gutzwiller-like projector can be understood as follows: consider an ``island'' of charge in the state $\ket{\cdots 0 11 \cdots 10 \cdots}$, where we have a string of $1$'s with length $\ell$. This ``island" contains $\ell-1$ nearest-neighbor occupied pairs, and so the projector gives a factor of $g^{\ell-1}$. The overall factor one obtains on any state (in the particle number basis) is the product of the factor from all of such islands of charges.
Because our MPS language is based on the fermionic system, we need to turn the projector into a fermionic matrix product form.
Its matrix product form is shown in Fig.\ \ref{fig:joint_gutzwiller_projector}.
The matrix can be written in a compact form:
\begin{equation}
        \hat{T}^{[x]} = \sum_{i_x,j_x}\vcenter{\hbox{\includegraphics[width = 0.1\textwidth]{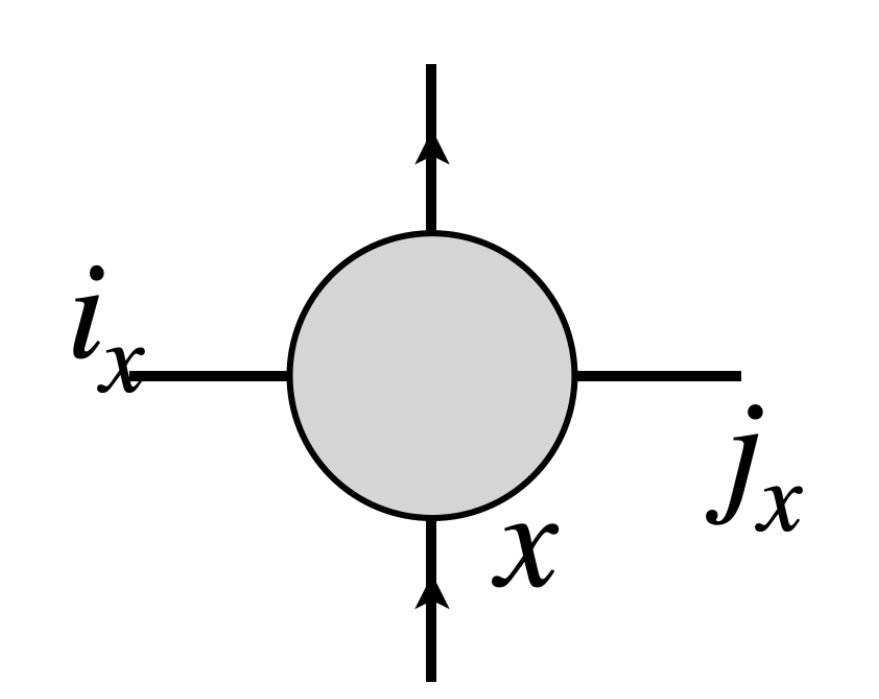}}} = \sum_{i_x,j_x}\ket{i_x} \hat{O}_{i_xj_x}^{[x]}\bra{j_x},
\end{equation}
where
\begin{equation}
    \hat{O}_{i_xj_x}^{[x]} = (a_L\hat{n}_x)^{i_x}(a_R\hat{n}_x)^{j_x};\quad a_La_R = g-1.
\end{equation}
where $i_xj_x\in \{0,1\}$. As a convenient choice, set $a_L = (g-1), a_R = 1$:

\begin{equation}
    \hat{O}_{00}^{[x]} = \hat{1};\quad \hat{O}_{01}^{[x]} = \hat{n}_x; \quad\&\quad\hat{O}_{10}^{[x]}=\hat{O}_{11}^{[x]} = (g-1)\hat{n}_x 
\end{equation}
Then a continuous chain of nearest-neighbor Gutzwiller projectors is recovered by the product of $\hat{T}^{[x]}$:
\begin{equation}
\begin{aligned}
    &\hat{T}^{[0]}\hat{T}^{[1]}\cdots\hat{T}^{[\ell]} \\
    =& \sum_{i_0,j_0,\cdots i_\ell,j_\ell}\ket{i_0}\hat{O}^{[0]}_{i_0j_0}\bra{j_0}i_1\rangle\hat{O}^{[1]}_{i_1j_1}\bra{j_1}i_2\rangle\cdots \hat{O}^{[\ell]}_{i_\ell j_\ell}\bra{j_\ell}\\
    =&\sum_{i_0,i_1,i_2\cdots i_{\ell},j_\ell} \ket{i_0}\hat{O}^{[0]}_{i_0i_1}\hat{O}^{[1]}_{i_1i_2}\cdots\hat{O}_{i_{\ell}j_\ell}\bra{j_\ell}\\
    =& \sum_{i_0,j_\ell} \ket{i_0}((g-1)\hat{n}_0)^{i_0}\\
    &\left[\sum_{i_1,i_2\cdots i_{\ell}}\prod_{n=1}^\ell(\hat{n}_{n-1})^{i_n}((g-1)\hat{n}_{n})^{i_n}\right](\hat{n}_\ell)^{j_\ell}\bra{j_\ell}\\
    =& \sum_{i_0,j_\ell} \ket{i_0}((g-1)\hat{n}_0)^{i_0}\left[\prod_{n=1}^\ell(1+(g-1)\hat{n}_{n-1}\hat{n}_{n})\right](\hat{n}_\ell)^{j_\ell}\bra{j_\ell}
\end{aligned}
\end{equation}
where the contraction of bond states satisfies
\begin{equation}
    \bra{j_x}i_{x+1}\rangle = \delta_{j_x,i_{x+1}}
\end{equation}
In fermionic form the purification of the operator $\hat{T^{[x]}}$ is a composition of the purified operator and bond states. The purified operator is defined as 
\begin{equation}
    \ket{\hat{O}_{i_xj_x}^{[x]}}:=\left(\hat{O}_{i_xj_x}^{[x]}\otimes\hat{1}\right)\ket{D^{[x]}},
\end{equation}
where the reference state $\ket{D^{[x]}}$ is defined on the Fock space of the physical fermion at x together with its auxiliary. Note that $D^{[x]}$ has charge 1 and is odd-parity. The same is true for $\hat{O}_{i_xj_x}^{[x]}   $, as all the $\hat{O}_{i_xj_x}^{[x]}$ are number conserving. The fermionic version of the bond states should satisfy
\begin{equation}
    \bra{K_{x,x+1}}\left(\ket{f_{j_x}}\otimes_f\ket{f_{i_{x+1}}}\right) = \bra{j_x}i_{x+1}\rangle = \delta_{j_x,i_{x+1}}
\end{equation}
To ensure the bond states follow the fermionic statistics, we use two fermions to represent the bond:
\begin{equation}
\begin{aligned}
    &\ket{f_{j_x=0}}=\hat{r}_{x;0}^\dagger\ket{0} = \ket{10}\\
    &\ket{f_{j_x=1}}=\hat{r}_{x;1}^\dagger\ket{0} = \ket{01}\\
    &\ket{f_{i_{x+1}=0}}=\hat{l}_{x;1}^\dagger\ket{0} = \ket{01}\\
    &\ket{f_{i_{x+1}=1}}=-\hat{l}_{x;0}^\dagger\ket{0} = -\ket{10}
\end{aligned}
\end{equation}
which matches our convention of contraction kernel
\begin{equation}
\begin{aligned}
    \ket{K} =& (\hat{r}_{x;0}^\dagger +\hat{l}_{x+1;0}^\dagger)(\hat{r}_{x;1}^\dagger + \hat{l}_{x+1;1}^\dagger)\ket{0}\\
    &(\hat{r}_{x;0}^\dagger\hat{r}_{x;1}^\dagger+\hat{r}_{x;0}^\dagger\hat{l}_{x+1;1}^\dagger + \hat{l}_{x+1;0}^\dagger\hat{r}_{x;1}^\dagger + \hat{l}_{x+1;0}^\dagger\hat{l}_{x+1;1}^\dagger)\ket{0}\\
    &\ket{1100}+\ket{1001}-\ket{0110}+\ket{0011}
\end{aligned}
\end{equation}
Combining the discussions above, we finally obtain the fermionic state corresponding to the local tensor
\begin{equation}
\begin{aligned}
    \ket{\hat{T}^{[x]}} =& \sum_{i_x,j_x = 0,1}\ket{f_{i_x}}\ket{\Omega_{i_xj_x}^{[x]}}\ket{f_{j_x}}\\
    =&\ket{01}\ket{\hat{O}_{00}^{[x]}}\ket{10}+\ket{01}\ket{\hat{O}_{01}^{[x]}}\ket{01}\\&-\ket{10}\ket{\hat{O}_{10}^{[x]}}\ket{10}-\ket{10}\ket{\hat{O}_{11}^{[x]}}\ket{01}
\end{aligned}
\end{equation}
\section{Contraction and Transfer matrix}\label{appendix:transfer_mat_and_computation}

In this section we first derive the formula Eq.\,\eqref{eq:wavefunction_contraction}, and then we show the derivation of Eq.\,\eqref{eq:trans_mat_exp_values}. For simplicity, we use the the vector symbol $\vec a^\dagger, \vec p^\dagger$ to denote the product of a sequence of creation operators, for instance, $\vec a_0^\dagger = (0,1,1)^\dagger \equiv \hat a_{0;2}^\dagger \hat a_{0;3}^\dagger $ representing the creation of mode 2 and 3 in the bond space labelled by $a_0$. The conjugate of these symbols $\vec a_0$ represent the conjugate of the creation operators $\hat a_{0;3}\hat a_{0;2}$.

\subsection{Contraction via matrix product }
To derive the Eq.\,\eqref{eq:wavefunction_contraction}, we consider a kernel contraction of local tensors,
\begin{equation}\begin{split}
        &\bra{K_{b,b'}}( \ket{\Psi_{a,b}}\otimes\ket{\Phi_{b',c}} ) \\
        = & K^*_{\vec b, \vec b' } \Psi_{\vec a \vec b} \Phi_{\vec b'\vec c}\bra{0} \vec b' \vec b \vec a^\dagger\vec b^\dagger \vec b^{\prime\dagger}\vec c^\dagger \ket{0}   \\
        = &K^*_{\vec b,\vec b'}\Psi_{\vec a\vec b}\Phi_{\vec b'\vec c} (-1)^{n_{\vec a}N_{b}} \vec a^\dagger\vec c^\dagger\ket{0} \\
        = & (P_{\vec a} )^{N_b} \Psi_{\vec a\vec b}K_{\vec b,\vec b'}\Phi_{\vec b'\vec c}{\vec a^\dagger\vec c^\dagger}\ket{0}.
\end{split}
\end{equation}

To show the kernel purification is compatible with kernel contraction, we need some properties of the contraction kernel $\ket{K}$ and purification kernel $\ket{D}$. The matrix product satisfies the following identity
\begin{equation}
    \sum_{\vec m'}D_{\vec m,\vec m'}K_{\vec m',\vec a} =(-1)^{n_{\vec a}(N_{\vec m})} = [\hat P]_{\vec m \vec a}^{N_{\vec m}}.
\end{equation}
Now consider an operator $\hat O_{m} = O_{\gamma}\hat \gamma$, $\hat\gamma$ being a product of some Majorana operators, acting on the modes $\vec m$. Then
\begin{equation}
\begin{split}
        & \sum_{\vec m'\vec a}\bra{K_{\vec m'\vec a}}\hat O_m \ket{D_{\vec m\vec m'}}\otimes_f\ket{\psi_{\vec a\vec b}}  \\
        = & K_{\vec m'\vec a} O_{\gamma} D_{\vec m\vec m'} \psi_{\vec a\vec b}\bra{0} \vec a \vec m' \gamma \vec{m}^\dagger \vec{m}'^\dagger \vec a^\dagger \vec b^\dagger \ket{0} \\
        = & O_{\gamma} D_{\vec m\vec m'}K_{\vec m'\vec a} \psi_{\vec a\vec b} (-1)^{N_\gamma (n_{\vec a}+n_{\vec m'})}\\
        & \times(-1)^{ n_{\vec m}(n_{\vec m'}+n_{\vec a}) } \ga \vec m^\dagger \vec b^\dagger\ket{0}, \\
\end{split}
\end{equation}
since $n_{\vec m}+n_{\vec m'} = N_m$ is fixed, and the only non-vaninshing elements in $DK$ are diagonal elements, thus $n_{\vec a} = n_{\vec m}$, the expression can be simplified to
\begin{equation}\begin{split}
        & (-1)^{N_\gamma N_m}O_\gamma \psi_{\vec a\vec b} (-1)^{n_{\vec a} (N_m)} (-1)^{n_{\vec m} N_m} \ga \vec m^\dagger \vec b^\dagger\ket{0} \\
        = & (-1)^{N_\gamma N_m}O_\gamma \psi_{\vec m\vec b} \ga \vec m^\dagger \vec b^\dagger \ket{0}
\end{split}
\end{equation}
So given the operator has an even parity, or the number of modes $N_{\vec m}$ it acts on is even, the result above is equal to the  $\hat O_{\vec m}\ket{\psi_{\vec m\vec b}}$. If $\hat O$ has an odd parity and $N_{\vec m}$ is odd, then it is up to a negative sign. For a generic $\hat O$ with terms of mixed parities, we can distinguish the parity sectors and compensate the signs accordingly.

\subsection{Transfer matrix}
The transfer matrix in this paper is defined as the regrouped matrix of a reduced density matrix tracing out the physical modes. Let $\ket{\phi_{L,P,R}}$ be a local fermion state with $N_L+N_P+N_R$ modes, the reduced density matrix is 
\begin{equation}
    \hat \rho_{LR} = \Tr_P\left[ \ket{\phi_{L,P,R}}\bra{\phi_{L,P,R}}\right]
\end{equation}
In the computational basis, it is a $2^{N_L N_R}\times 2^{N_L N_R}$ matrix. Then regrouping the elements leads to a transfer matrix $M_{ll', rr'} = (\hat \rho_{LR})_{lr,l'r'}$. This definition is different from the usual transfer matrix defined in the literature, as the partial trace of a fermionic density matrix involves some sign differences. 
We now show the computation of observables can be represented as computing the trace of a product of transfer matrices, without the need of inserting the parity operators $\hat P^{N_B}$ in Eq.\,\eqref{eq:wavefunction_contraction}. Suppose there are $L$ local tensors $\ket{\Phi_{a_0 p_1 a_1} }, \ket{\Phi_{\tilde a_1 p_2 a_2}}, ... \ket{\Phi_{\tilde a_{L-1} p_L a_L}}$ and denote their tensor product as $\ket{\Phi} = \ket{\Phi_{a_0 p_1 a_1} }\otimes\ket{\Phi_{\tilde a_1 p_2 a_2}}\otimes...\otimes\ket{\Phi_{\tilde a_{L-1} p_L a_L}}$, we first consider the computation of normalization
\begin{equation}\label{eq:compute_normalization}
    \mathcal{N}_{\Phi} = \ev{\Phi|K}\ev{K|\Phi} = \Tr\left[ \ev{K|\Phi}\ev{\Phi|K}\right],
\end{equation}
where $\ket{K} = \ket{K_{a_1\tilde a_1}}\otimes\ket{K_{a_2\tilde a_2}}\otimes...\otimes \ket{K_{a_L a_0}}$ is the product state of all contraction kernels. The states $\ket{K}$ and $\ket{\Phi}$ are denoted as 
\begin{equation}\begin{split}
        \ket{K} &= (K_{\vec a_0 ... }) \vec a_1^\dagger \tilde{\vec a}_1^\dagger \vec a_2^\dagger \tilde{\vec a}_2^\dagger\cdots\vec a_{L}^\dagger\vec{a}_0^\dagger \ket{0}; \\
        \ket{\Phi} &=(\Phi_{\vec a_0...}) \vec a_0^\dagger \vec p_1^\dagger \vec{a}_1^\dagger... \tilde{\vec a}_{L-1}^\dagger \vec p_{L}^\dagger \vec{a}_{L}^\dagger\ket{0},
\end{split}
\end{equation}
where the coefficients are also labelled by these vectors. The Eq.\,\eqref{eq:compute_normalization} is done by first obtaining the local reduced density matrices and then trace with the kernel state,
\begin{equation}
\begin{split}
        \mathcal{N}_{\phi} &= \bra{K} \Tr_P\left[\ev{\Phi|\Phi}  \right] \ket{K} \\
        & = K_{\vec a_0...}K^*_{\vec a'_0...} M^{\vec a_0...}_{\vec a'_0...} \bra{0} \underbrace{\vec{a}_0\vec{a}_L...\tilde{\vec{a}}_1\vec{a}_1}_{K}  \underbrace{\vec{a}_0^\dagger\vec{a}_1^\dagger\tilde{\vec{a}}_1^\dagger... \vec{a}_L^\dagger}_{\Phi \text{ reduced}} \ket{0} \\
        & \times \bra{0} \underbrace{\vec{a}_L ...\tilde{\vec{a}}_1  \vec{a}_1  \vec{a'}_0   }_{\Phi \text{ reduced}} \underbrace{ \tilde{\vec{a}}_1^{\prime \dagger} \vec{a'}_1^\dagger ...\vec{a'}_L^\dagger \vec{a'}_0^\dagger }_{K} \ket{0},
\end{split}
\end{equation}
with all repeated indices summed up implicitly. 
Now the only possibly non-trivial signs coming from these creation and annihilation operators are those due to the $\vec a_0/\vec a_0^\dagger$ and $\vec a_L/\vec a_L^\dagger$, which is  
\begin{equation}
\begin{split}
    &(-1)^{n_{\vec a_0}(N_{\vec a_1}+...+N_{\vec a_{L-1} }) + n_{\vec a_0}n_{\vec a_L} }\\
    &\qquad \times (-1)^{n_{\vec a'_0}(N_{\vec a'_1}+...+N_{\vec a'_{L-1} }) + n_{\vec a'_0}n_{\vec a'_L} }
\end{split}
\end{equation}
where we used the condition that each kernel state is particle-number-conserving. A further simplification is $n_{\vec a_0}\equiv n_{\vec a_L}+ N_{\vec a_0} \text{mod 2}$, so the above sign is
\begin{equation}
    (-1)^{(n_{\vec a_0}+n_{\vec a'_0})(N^{\vec t}_{\vec a}+1) },
\end{equation}
where $N^{\vec t}_{\vec a}$ is one half of the total number of bond modes of $\ket{\Phi}$. So the computation of the normalization is reduced to 
\begin{equation}
\begin{split}
        \mathcal{N}_{\Phi} &= M^{\vec a_0 \vec a_1}_{\vec a'_0, \vec a'_1}\mathcal{K}^{\vec a_1,\tilde{\vec a}_1 }_{\vec a'_1,\tilde{\vec a}'_1}\cdots M^{\tvec{a}_{L-1},\vec a_L}_{\tvec{a}'_{L-1},\vec a'_L} \mathcal{K}^{\vec a_L,\vec a_0}_{\vec a'_L,\vec a_0'} \left[\Xi^{N^{\vec t}_{\vec a} +1}\right]^{\vec a_0}_{\vec a'_0} \\
        &= \text{tr}\left[ M_0 \mathcal{K}_{0,1} M_1 \mathcal{K}_{1,2}\cdots M_L\mathcal{K}_{L,0} \Xi^{N^{\vec t}_\vec a+1} \right], 
\end{split}
\end{equation}
where we use $\mathcal{K} = K\otimes K^*$ to denote the Kronecker product of the matrices $K_{\vec a_i, \tvec{a}_i} $ and $K^*_{\vec a'_i, \tvec{a}'_i} $, and $\Xi$ to denote the Kronecker product of the parity matrices $\hat P_{\vec a_0, \vec a_0}$ and $\hat P_{\vec a'_0,\vec a'_0}$.
The same derivation can be done for the computation of observables. With the insertion of some operators, the coefficients of $\Phi$ will be modified, but as long as the transfer matrix is defined as the partial trace of physical modes in local tensors, the sign issue is always the same as derived above.

\section{Finite-bond scaling}\label{appendix:finite_bond_scaling}
In the main text we discussed the tree-stack construction of translation-invariant local tensors. This process involves finding a fGMPS representation, which includes a radius $r$ truncation around the Wannier center. The truncation controls the number of bonds of local tensor, affecting the momentum distribution $n_k$ near the Fermi surface. We define $\Delta k$ to describe the smearing:
\begin{equation}
    \Delta k(\epsilon):=\frac{N_\epsilon(n_k)}{L}
\end{equation}
where $N_\epsilon(n_k)$ is the range of momentum $k$ satisfying $\epsilon<n_k<1-\epsilon$ and $L$ is the system size.
\begin{figure}[htpb]
    \centering
    \includegraphics[width=0.5\textwidth,left]{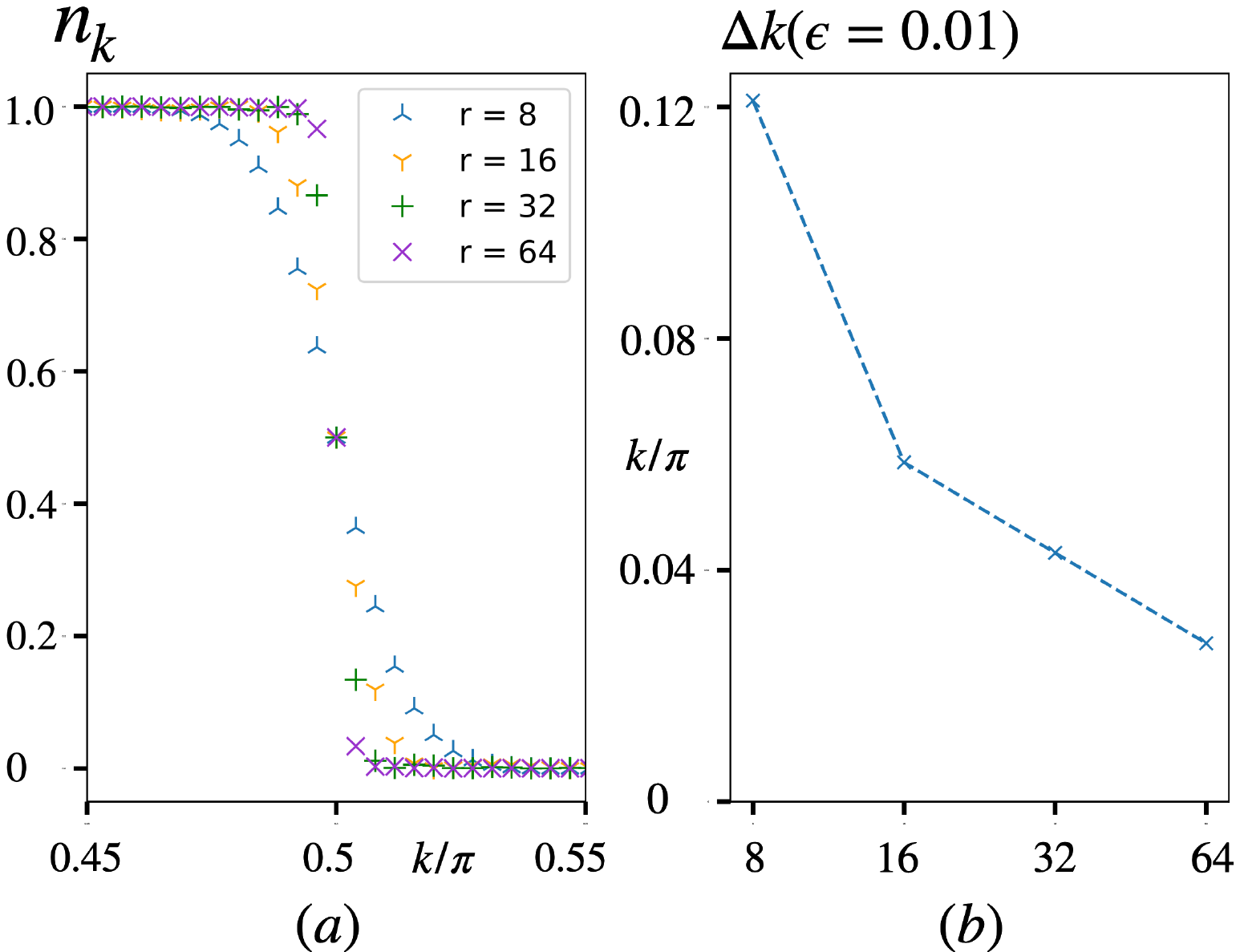}
    \caption{The computation here is based on a half-filled Fermi sea of spin-$1/2$ electrons in one dimension system. (a)
    The momentum density $n_k$ distribution performance near $\pi/2$ for different stack tree radius. (b) The $\Delta k (\epsilon = 0.01)$ for truncation radius $r = 8, 16, 32, 64$ for a system with system size $L =512 $.}
    \label{fig:n_delat_k_scaling}
\end{figure}
A smaller $\Delta k(\epsilon)$ represents a sharper Fermi surface.\\

In Fig.\ \ref{fig:n_delat_k_scaling}, we show the momentum density distribution and $\Delta k$ for different truncation length in a system with size $L=512$ model. As the truncation radius $r$ increase, the momentum density $n_k$ near $k = \pi/2$ moves closer to $n_k = 0,1$, resulting in a decrease in $\Delta_k(\epsilon = 0.01)$ . Ultimately, in the thermodynamic limit $\Delta k$ is expected to vanish.

\newpage

\bibliography{references}
\clearpage

\end{document}